\documentclass[journal=jpcbfk,manuscript=article]{achemso}
\usepackage{times}
\usepackage{graphicx}
\usepackage{psfrag}
\usepackage{epsfig}
\usepackage{ae}
\usepackage{amsmath,amssymb}
\usepackage{booktabs}
\usepackage[version=3]{mhchem} % Formula subscripts using \ce{}
\usepackage[usenames]{color}
\usepackage{color,soul}
\usepackage{float}
\usepackage{subfigure}
\author{T. P. O. Nogueira} 
\email{thiagoponogueira@gmail.com}
\affiliation{Departamento de F\'{i}sica, Instituto de F\'{i}sica e Matem\'{a}tica, 
    Universidade Federal de
    Pelotas. Caixa Postal 354, 96001-970, 
    Pelotas, Brazil.}
    
\author{H. O. Frota} 
\affiliation{Department of Physics, 
    Federal University of Amazonas, 
    69077-000 Manaus, AM, Brazil.}    

\author{Francesco Piazza} 
\email{Francesco.Piazza@cnrs-orleans.fr}
\affiliation{Universit\'{e} d'Orl\'{e}ans, Centre de Biophysique 
    Mol\'eculaire (CBM), CNRS UPR4301, 
    Rue C. Sadron, 45071 Orl\'{e}ans, France.}

\author{Jos\'e Rafael Bordin} 
\email{jrbordin@ufpel.edu.br}
\affiliation{Departamento de F\'{i}sica, Instituto de F\'{i}sica e Matem\'{a}tica, 
    Universidade Federal de
    Pelotas. Caixa Postal 354, 96001-970, 
    Pelotas, Brazil.}

\title{Tracer diffusion in crowded solutions of sticky polymers}

 \date{\today}

\begin{document}

%  \begin{tocentry}
%  \includegraphics[width=6cm]{cover.png}
%  \end{tocentry}

\begin{abstract}

        Macromolecular diffusion in strongly confined geometries and crowded environments
        is still to a large extent an open subject in soft matter physics and biology. 
        In this paper, we employ large-scale Langevin dynamics simulations 
        to investigate how the diffusion of a tracer is influenced by the combined 
        action of excluded-volume and weak attractive crowder-tracer interactions.
        We consider two species of tracers, standard hard-core particles described by the
        Weeks-Chandler-Andersen (WCA) repulsive potential and
        core-softened (CS) particles, 
        which model, e.g., globular proteins, charged colloids and nanoparticles covered by 
        polymeric brushes. These systems are characterized by the 
        presence of two length scales in the interaction and 
        can show water-like anomalies in their diffusion, stemming from the inherent 
        competition between different length scales. 
        Here we report a comprehensive study of both 
        diffusion and structure of these two tracer species in an environment crowded by 
        quenched configurations of polymers at increasing density. 
        We analyze in detail how the tracer-polymer affinity and the system density 
        affect  transport as compared to the emergence of specific static spatial correlations.
        In particular, we find that, while hardly any differences emerge in the diffusion 
        properties of WCA and CS particles, the propensity to develop structural order 
        for large crowding is strongly frustrated for CS particles.
        Surprisingly, for large enough affinity for the crowding matrix, 
        the diffusion coefficient of WCA tracers display a non-monotonic trend as 
        their density 
        is increased when compared to the zero affinity scenario. This water-like anomaly 
        turns out to be even larger than what observed for CS particle and appears 
        to be rooted in a similar competition between excluded-volume and affinity effects.

\end{abstract}

% \pacs{64.70.Pf, 82.70.Dd, 83.10.Rs, 61.20.Ja}

\maketitle
\section{Introduction}

    Diffusion process are well described by conventional fluid mechanics for
    dilute mixtures in non-complex geometries. This is not the case
    in biological environments, such as the cell, which contains a high diversity of 
    macromolecular species at high concentrations in strongly confining geometries. 
    For instance, in a cell macromolecules such as proteins, 
    polysaccharides, nucleic acids and other smaller molecules
    occupy volume fractions typically larger than 30\% of the available space
    ~\cite{zimmerman1991estimation,zimmerman1993macromolecular,zhou2004protein,
    zhou2008macromolecular,mcguffee2010diffusion}.\\
    \indent While excluded-volume effects in such contexts are often referred 
    to as {\em macromolecular crowding}~\cite{rivas2016macromolecular}, 
    other factors strongly affect diffusion-based processes 
    in biological media, most importantly many non-specific interactions, such as 
    van de Waals, electrostatic and hydrodynamic 
    interactions~\cite{blanco2018macromolecular}. 
    Indeed, crowded environments may influence  the behavior of biomolecules in many subtle ways.
    For example, it has been proved that stabilization or destabilization 
    of proteins and protein complexes may be observed, depending on the relative strength 
    of weak non-specific enthalpic interactions
    and purely entropic excluded-volume interactions~\cite{bhattacharya2013}.  \\  
    %
    %For instance, it 
    %was found that the macromolecular crowding promotes self-association of FtsZ, 
    %accelerating the rate of amyloid formation~\cite{rivas2000magnesium,hatters2002macromolecular}. 
    %Specially, it substantially alters the diffusion processes occurring inside these 
    %environments~\cite{ellis2001macromolecular,balcells2014macromolecular}.
    %
    \indent Overall, the influence of crowding and non-specific interactions on 
    single-particle diffusion are not fully understood~\cite{rivas2016macromolecular}.
    While single-particle tracking (SPT) experiments and fluorescence correlation 
    spectroscopy (FCS) allow one in principle  to
    measure the diffusion coefficient of fluorescently tagged tracers {\em in vivo}~\cite{dix2008crowding}, 
    the interpretation of the results in the context of simple models is still a difficult
    task~\cite{blanco2018macromolecular}.
    One of the easiest and most interesting workarounds to this problem is to measure 
    diffusion in controlled artificial environments, created so as to mimic only selected specific 
    features  of the overwhelming cellular complexity. For example, tunable crowded environments
    where non-specific interactions are virtually silenced have been realized using highly concentrated 
    polymer solutions, such as dextran or ficoll~\cite{banks2005anomalous,pastor2010diffusion},
    where diffusion has been investigated, e.g., through FCS~\cite{banks2005anomalous} and 
    fluorescence recovery after photobleaching~(FRAP) spectroscopy~\cite{pastor2010diffusion}. \\
    \indent In addition to the above mentioned and other experimental studies,
    numerical simulations have proved invaluable to dissect the subtle influence of the 
    environment on the mobility of tracers~\cite{blanco2018macromolecular}.
    Many different computational techniques  have been employed to this end, from Monte Carlo 
    simulations~\cite{vilaseca2011diffusion,vilaseca2011new} and
    Brownian Dynamics~(BD)~\cite{sun2007toward,mereghetti2012atomic,kondrat2015effect,
    sentjabrskaja2016anomalous,yu2016biomolecular,blanco2017brownian,smith2017fast}
    to Molecular Dynamics simulation~\cite{yu2016biomolecular,bucciarelli2016dramatic,wang2017influence,chakrabarti2013tracer}.\\
    \indent As a general rule, biological macromolecules tend to have variably flexible structures. Therefore, 
    when two macromolecules approach each other, their branches can become entangled 
    and non-specific attractive and/or repulsive interactions may come into play. This effect 
    is expected to influence to a substantial extent 
    macromolecular diffusion in crowded media. This, in turn, implies that 
    models based exclusively on hard spheres may miss important features 
    of the tracer-environment interactions~\cite{irback2017protein}.\\
    \indent In the simplest approximation beyond hard-core interactions, 
    core-softened colloids and proteins may be characterized by the presence 
    of two length scales,  a short-range attraction
    and a long-range repulsion~\cite{Stradner04, Shukla08}.
    The repulsion can be caused by a soft shell, as in the case of polymeric 
    brushes~\cite{Lafitte14, Curk14, Nie16,bedrov2006multiscale,bedrov2007structure,samanta2016tracer,kumar2019transport}
    and star-polymers~\cite{likos1998star,loppinet2005dynamics,liu2014coarse,locatelli2016multiblob}, or
    by electrostatic repulsion in charged colloids, macromolecules, lysozyme and spherical 
    proteins~\cite{vilaseca2011diffusion, Shukla08, QP01, CA10}, while the 
    attraction is caused by van der Waals forces or solvent effects~\cite{Alamarza14, bordin2018b}.\
    \indent Core-softened potentials have also been largely employed to study systems with 
    water-like anomalies~\cite{vilaseca2011diffusion, barros2006thermodynamic,de2006structural, bordin2018}
    along with the related confinement effects~\cite{BoK15b, Krott13b, BoK15a}.
    In fact, for most materials the diffusion coefficient decreases when the pressure (or density)
    increases. However, anomalous materials such as 
    water~\cite{Ne02a}, silicon~\cite{Mo05} and silica~\cite{Sa03} 
    also display a diffusion anomaly, characterized by a maximum in the diffusion 
    coefficient at constant temperature.
    Therefore, a relevant question that arises is how  crowded media may influence the diffusion of
    core-softened tracer particles immersed in a polymeric environment with respect 
    to hard-core particles.
    To answer this question, in this paper we employ large-scale Langevin dynamics simulations 
    and characterize how excluded-volume effects and tracer-crowder interactions 
    affect tracer diffusion. More precisely, we consider two species of tracers, 
    (i) hard-core spherical particles modeled by the well known 
    Weeks-Chandler-Andersen~(WCA) purely repulsive potential~\cite{weeks1971}, and (ii) 
    core-softened particles, displaying a dual-length scale hard core--soft corona structure. 
    In order to mimic  macromolecular crowding,
    we inject these tracer molecules in a complex static polymeric solution realized 
    via a coarse-grained model~\cite{Kremer1990}. Our main goal is to investigate 
    single-particle diffusion and static space correlation of these two molecular species 
    immersed in the polymer matrix.
    More precisely,  we aim at 
    analyzing how the tracer-polymer affinity and the polymer density affect the 
    transport and spatial correlations of the tracer particles. 
    
%%%%%%%%%%%%%%%%%%%%%%%%%%%%%%%%%%%%%%%%%%%%%%%%%%%%%%%%%%%%%%%%%%%%%%%%%%%%%%%%%%%%%%%%%%%%%%%%%%%%%%%%%%%
\section{Model and Simulation Details}
%%%%%%%%%%%%%%%%%%%%%%%%%%%%%%%%%%%%%%%%%%%%%%%%%%%%%%%%%%%%%%%%%%%%%%%%%%%%%%%%%%%%%%%%%%%%%%%%%%%%%%%%%%%

%%%%%%%%%%%%%%%%%%%%%%%%%%%%%%%%%%%%%%%%%%%%%%%%%%%%%%%%%%%%%%%%%%%%%%%%%%%%%%%%%%%%%%%%%%%%%%%%%    
    \setcounter{subfigure}{0}% Reset subfigure counter  
    \begin{figure}[h!]
        \centering
        \subfigure[]{\includegraphics[width=0.4\textwidth]{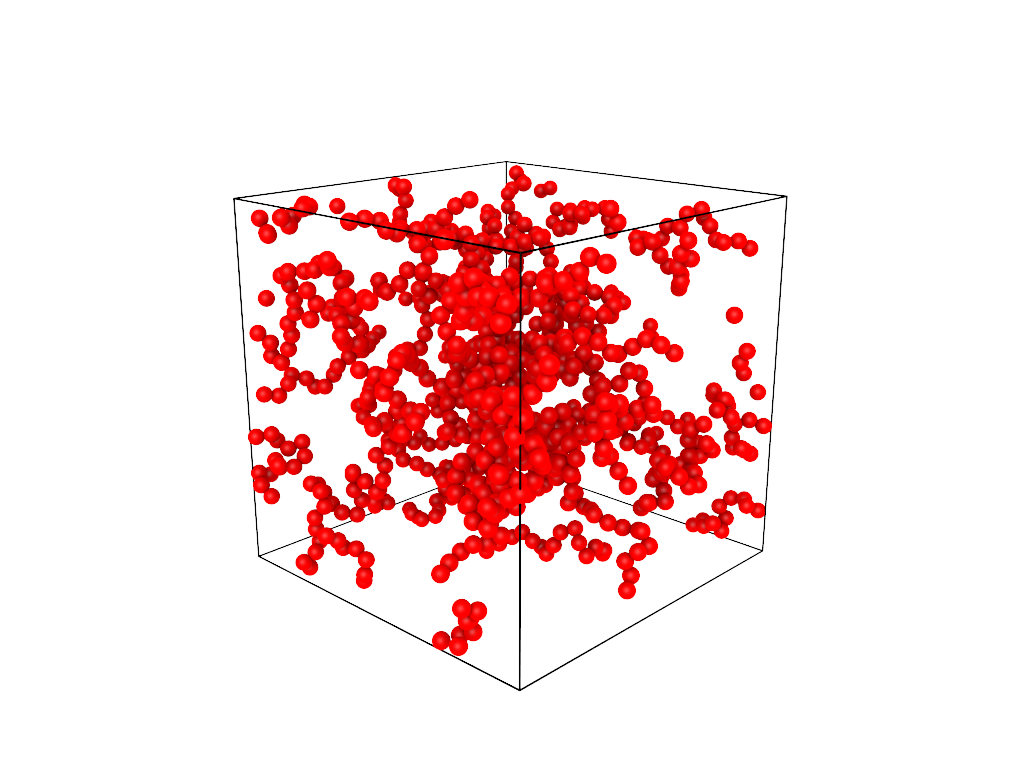}}
        \subfigure[]{\includegraphics[width=0.4\textwidth]{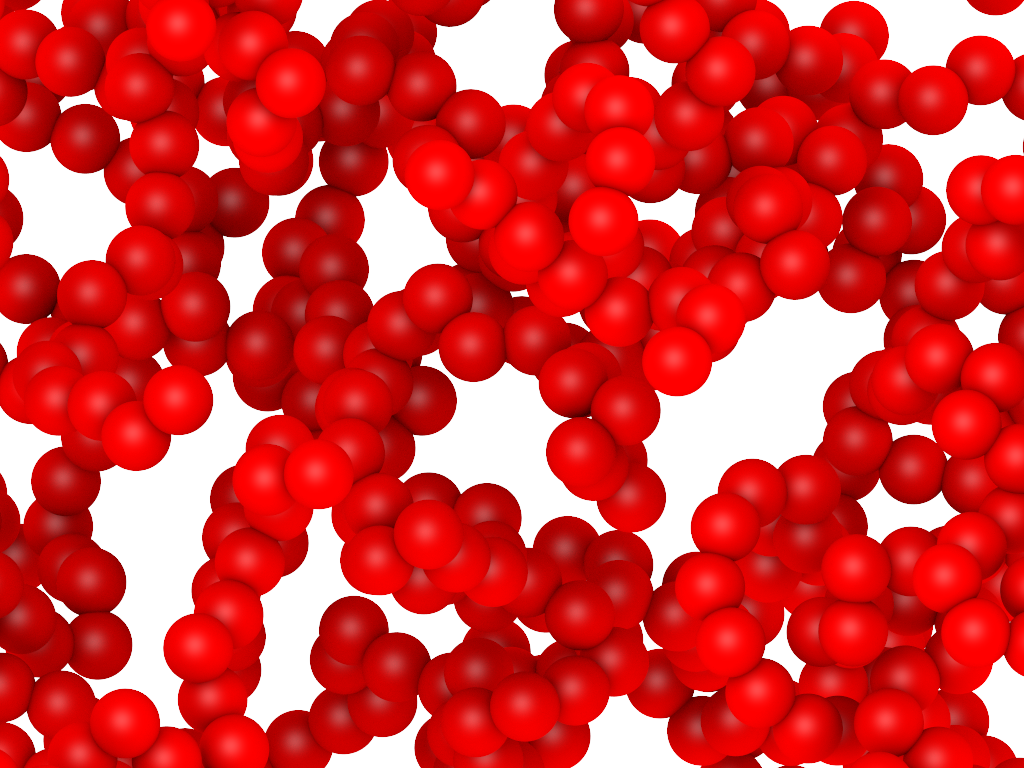}}
        \subfigure[]{\includegraphics[width=0.4\textwidth]{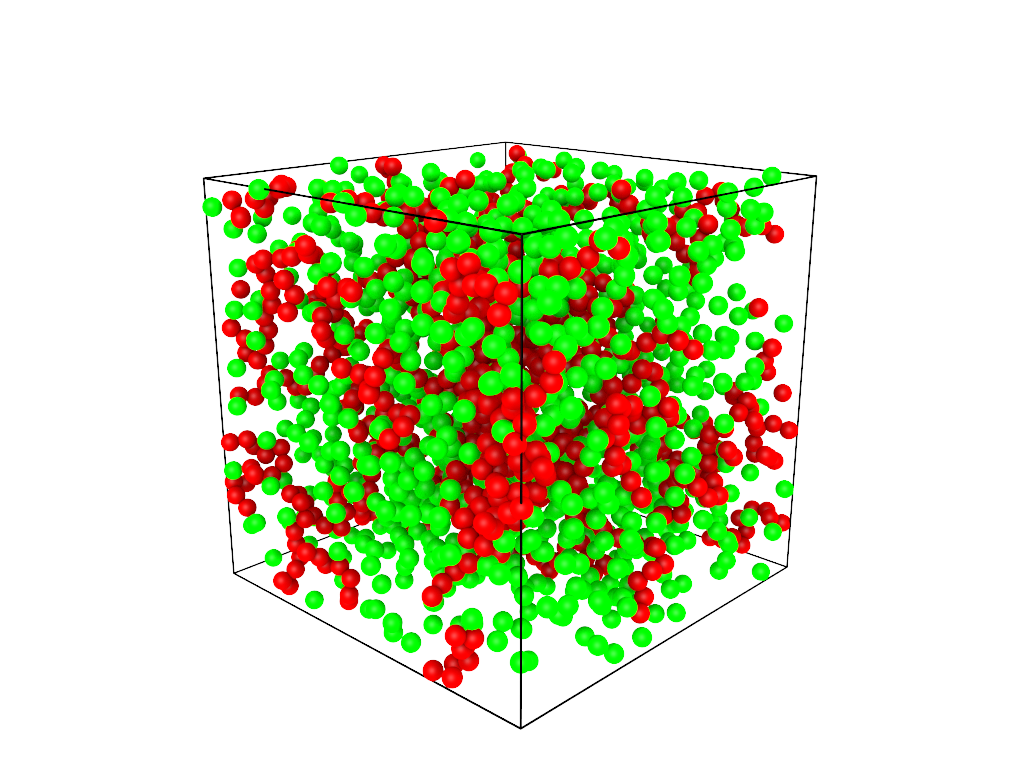}}
        \caption{(Color online) Illustration of the three-dimensional 
        tracer-polymer system analyzed in our 
        simulations. (a) System snapshot with only polymers obstacles at volume occupancy
        $\phi_P = 0.05$. (b) Zoomed snapshot with only polymers obstacles at the same 
        volume fraction as seen in (a), where  a complex structure of voids 
        can be appreciated. 
        (c) Snapshot of the whole system, comprising the polymer obstacles (red) and the 
        tracer particles (green).}
        \label{fig1}
    \end{figure}
%%%%%%%%%%%%%%%%%%%%%%%%%%%%%%%%%%%%%%%%%%%%%%%%%%%%%%%%%%%%%%%%%%%%%%%%%%%%%%%%%%%%%%%%%%%%%%%%%  

\begin{figure}[h!]
        \centering
        \includegraphics[width=0.6\textwidth]{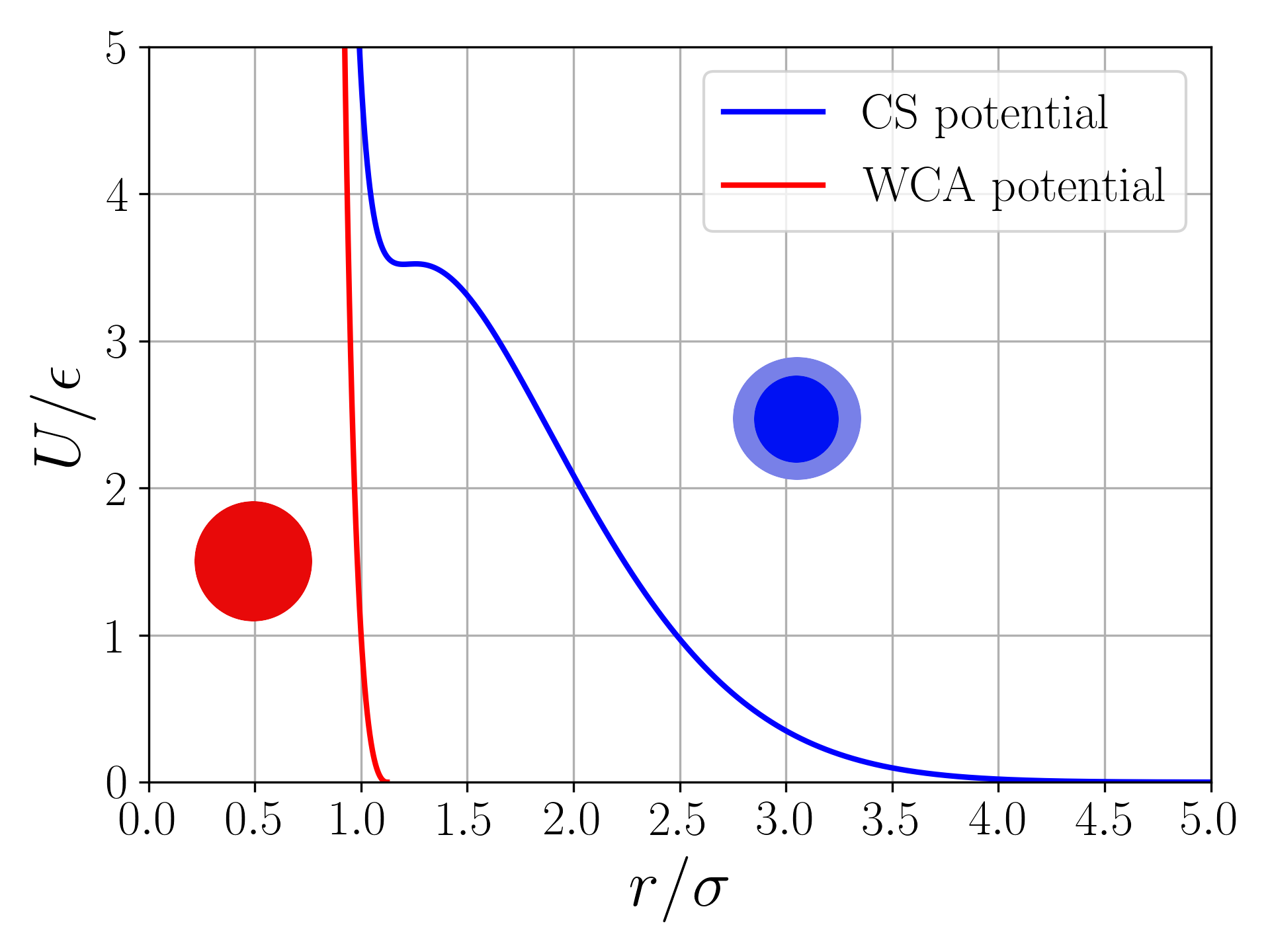}
        \caption{(Color online) Weeks-Chandler-Andersen~(WCA) and Core-softened~(CS)
        interaction potentials between two tracer particles. Inset: schematic 
        depiction of the particles, in red the WCA particle with its core
        (first length scale at $r=1.2\sigma$), in blue the CS 
        particle with its core (first length scale at $r=1.2\sigma$) and the 
        soft corona (second length scale at $r = 2.0 \sigma$). }
        \label{fig2}
    \end{figure}

    A fluid of tracer molecules with packing fraction $\phi_f$ 
    is immersed in a static polymer matrix such as the one 
    shown in Fig.~\ref{fig1}. We consider two types of tracer-tracer interactions.
    Single-length repulsive tracers interact through the WCA potential, $U_{\text{WCA}}$, namely
    \begin{equation}
        U_{\text{WCA}}(r) =\left\{ \begin{aligned}
                    & 4\epsilon \bigg[\bigg(\frac{\sigma}{r}\bigg)^{12} - 
                    \bigg(\frac{\sigma}{r}\bigg)^{6} + \frac{1}{4} \bigg] , \ r  \leqslant r_c  \\
                    & 0,\qquad\qquad \ r\geqslant r_c.
                    \end{aligned} \right.
    \label{eq:WCA}
    \end{equation}	
    The WCA potential is a standard potential energy function in soft matter, which was originally 
proposed to model a soft-core, purely repulsive potential~\cite{weeks1971}. 
It is obtained by cutting a standard 12-6 LJ potential  
at the position of the minimum, $r_c = 2^{1/6}\sigma$, and shifting this function upwards by
an energy equal to the well depth.  
As a consequence, the potential is zero above $r_c$ and has a sharp 
increase to $+\infty$ below $r_c$. The resulting potential is the soft-core purely repulsive red 
curve in Fig.~\ref{fig2}.\\
    \indent  Two-length scales repulsive tracers interact through 
    a core-softened interaction $U_{\text{CS}}$, 
    \begin{equation}
        U_{\text{CS}}(r) = 4\epsilon \bigg[\bigg(\frac{\sigma}{r}\bigg)^{12} - 
                           \bigg(\frac{\sigma}{r}\bigg)^{6}  \bigg] +
        u_0 \exp\bigg[-\frac{1}{c_{0}^{2}}\bigg(\frac{r-r_0}{\sigma}\bigg)^2\bigg].
        \label{eq:CS}
    \end{equation}
   \noindent The potential~\eqref{eq:CS} consists of the sum of a short-range Lennard-Jones term and a 
   long-range Gaussian function centered at $r_0$, with depth $u_0$ and width $c_0$. 
   The potential~\eqref{eq:CS} can be parameterized to have distinct shapes~\cite{barros2006thermodynamic,bordin2018b}. 
   Here we use the parameters $u_0 = 5 \epsilon$, $c_0^2 = 2.0$, and $r_0 / \sigma = 0.7$, 
   which lead to the repulsive ramp  depicted by the blue curve in Fig.~\ref{fig2}. 
   This shape was extensively applied to study systems with water-like anomalies~\cite{barros2006thermodynamic,
   de2006structural,bordin2018}, in view of its two characteristic length scales.  
   The first one is $r_1 = 1.2 \sigma$, 
   where the force has a local minimum, while the longer length scale is $r_2 = 2\sigma$,  
   where the fraction of imaginary modes of the instantaneous normal modes spectrum has 
   a local minimum~\cite{barros2010entropy}. 
   The cutoff radius for this interaction was fixed as $r_c^{CS} = 3.5 \sigma$.
%
%%%%%%%%%%%%%%%%%%%%%%%%%%%%%%%%%%%%%%%%%%%%%%%%%%%%%%%%%%%%%%%%%%%%%%%%%%%%%%%%%%%%%%%%%%%%%%%%%%%%%%%%%%     
    \par 
%%%%%%%%%%%%%%%%%%%%%%%%%%%%%%%%%%%%%%%%%%%%%%%%%%%%%%%%%%%%%%%%%%%%%%%%%%%%%%%%%%%%%%%%%%%%%%%%%%%%%%%%%%    
%
    \indent Five different polymer matrices were generated for each point 
    in parameter space by extracting independent configurations 
from $NVT$ equilibrium molecular dynamics of $N_c$ polymer chains consisting of an equal 
number $N_m$ of monomers of the same size $\sigma$ each, interacting along the polymer backbone 
through a standard 
Finitely Extensible Nonlinear Elastic (FENE) potential~\cite{Kremer1990,Auhl2003},
    \begin{equation}
        U_{\text{FENE}}(r) =\left\{ \begin{aligned}
                            & -\frac{1}{2}kR_0^2\ln\bigg[1 - \bigg(\frac{r}{R_0}\bigg)^2\bigg] , 
                            \ r  \leqslant R_0  \\
                            & \infty,\qquad\qquad \ r\geqslant R_0,
                            \end{aligned} \right.
        \label{eq:FENE}
    \end{equation}
    Here  $k$ is the elastic constant and $R_0$ is the limit distance for bond elongation. 
    In this work
    we used the same parameters  as in Ref.~\cite{Kremer1990}, namely 
    $k=30\epsilon/\sigma^2$ and $R_0=1.5\sigma$.
    Self-avoidance of the polymer chains was enforced by means of a Weeks-Chandler-Andersen (WCA)
    potential, acting for simplicity among all monomer pairs. Once the required number of
    independent polymer matrices were 
    generated, these were frozen and the fluid tracer  
    molecules were injected in the simulation box and the production runs started,
    the tracers being the only mobile species in all the simulations. \\
    \indent The interaction between tracers and the polymer matrix was modeled as a  
    standard LJ interaction
    \begin{equation}
        U_{\text{LJ}}(r) =\left\{ \begin{aligned}
                        & 4\epsilon_{\text{AB}} \bigg[\bigg(\frac{\sigma}{r}\bigg)^{12} - 
                        \bigg(\frac{\sigma}{r}\bigg)^{6}  \bigg] , \ r  \leqslant r_c  \\
                        & 0,\qquad\qquad \ r\geqslant r_c,
                        \end{aligned} \right.
        \label{eq:LJ}
    \end{equation}
    with cutoff  $r_c^{LJ} = 2.5\sigma$. 
     The tracer-polymer affinity $\epsilon_{\text{AB}}$  (A for polymer and B for 
    tracers) is one of the main parameters whose 
    effect on the tracer diffusion we want to investigate, with the aim of clarifying
    the competition between tracer-tracer and tracer polymer interactions. \\
    \indent All the simulations (generation of the polymer configurations and production runs) 
    were carried out in the $NVT$ ensemble, enforced by means 
    of Langevin dynamics (LD)~\cite{allen2017}
    in reduced LJ units, namely
    \begin{equation}
        r^{\ast} \equiv \frac{r}{\sigma}, \quad 
        \rho^{\ast} \equiv \rho\sigma^{3}, \quad
        \ t^{\ast} \equiv t \bigg(\frac{\epsilon}{m\sigma^2} \bigg)^{1/2}, \quad  
        T^{\ast} \equiv\frac{k_B T}{\epsilon},
    \end{equation}
    where $m$ is the mass of tracers and monomers. 
    The temperature was kept 
    constant via the Langevin thermostat at $T^\ast = 4.45$, which  
    ensured that the system was fluid even at high densities for both tracer species.    
    The packing fraction is defined as $\phi = N v / L^3$, where 
    $L$ is the side of the (cubic) simulation box, $v = \pi \sigma^3/6$ 
    is the single-particle volume and $N = N_t + N_mN_c$ is the total number 
    of particles, comprising $N_t$ tracers and $N_mN_c$ monomers belonging to the ensemble of polymer chains. 
    In practice, we fixed $N_c=100$ and $N_m=40$ in all simulations and varied $L$ and 
    $N_m$ in order to fix the fluid and polymer packing fractions,
    $\phi_f$ and $\phi_P$, at the chosen values.
    In the first set of simulations, we fixed the fluid packing fraction $\phi_f = 0.1$ and 
    varied the polymer packing fraction $\phi_P$ from 0.001 up to 0.1. In a second 
    set of simulations, we kept the polymer volume fraction fixed instead at 
    $\phi_P = 0.1$ and varied $\phi_f$ from 0.1 up to 0.4 with the 
    aim of investigating the effects of self-crowding.
    The time step used in the simulations was $\delta t = 0.001$ (reduced units), 
    and periodic boundary 
    conditions were enforced in the three Cartesian directions. Each simulation
    comprised a first relaxation and equilibration run ($10^6$ steps) 
    followed by $10^7$ steps 
    for the  production runs. To ensure that the system had reached equilibrium, 
    the kinetic and potential energy were monitored as functions of time. 
    All simulations 
    were carried out using the LAMMPS simulation package~\cite{plimpton1995fast}.\\
    \indent Five independent simulations were performed at each point in parameter space, 
    starting from as many independent random configurations of tracers and polymers.
    The diffusion coefficient was computed by fitting the long-time linear trend of the
     mean square displacement (MSD) as function of time, namely 
      \begin{equation}
        D = \lim_{t \rightarrow \infty} \frac{\mu_2(t)}{6t}\;.
    \end{equation}
    where the MSD $\mu_2(t) = \langle\langle [\vec r(t+t_0) - \vec r(t_0)]^2 \rangle\rangle$
    at a given lag $t$ was computed by the usual double average over the five independent
    runs and different time origins $t_0$.  The fluid structure was analyzed by computing  
    the radial distribution function (RDF) $g(r)$. More precisely, we evaluated 
    the RDF between particle A (polymer) and B (tracer) $g_{\text{AB}}(r)$ and the
    tracer-tracer RDF, $g_{\text{BB}}(r)$. \\
    \indent We point out that in this work hydrodynamic interactions (HI) are neglected,
    as the fluid is treated implicitly through the simple Langevin  scheme.  
    Indeed, HI have been shown to be 
important in accounting for the observed reduction of translational 
diffusion coefficient in very crowded, heterogeneous cell-like systems,
i.e. for densities around 300 mg/mL~\cite{Ando2010}. However, HI effects 
are expected to be prominent for non-spherical systems and in the presence of a wide 
range of molecular sizes in the system. In particular, one may expect that 
inclusion of HI would become important when the focus is on how crowding and molecular size
shapes diffusion for strongly non-spherical, heterogeneous systems.
In the case of the present study, we concentrate on rather homogenous 
spherical systems at moderate crowding ($\phi_P \leq 0.1$), 
where the radius of tracers is not varied. Therefore,
we expect that HI should not introduce a substantial bias in the {\em normalized}
diffusion coefficients, the main observable investigated in this work.\\
    \indent As a final remark, we note that in the following 
    all physical quantities are expressed in reduced LJ units,  
    while we will omit the superscript $^\ast$ for the sake of clarity.

%++++++++++++++++++++++++++++++++++++++++++++++++++++++++++++++++++++++++++++++++++++++++++++++++++++++++++
    \section{Results and Discussion}
    
    The first set of simulations refers to a fluid of tracers whose volume fraction was
    fixed at $\phi_f = 0.1$ and immersed in a polymer matrix that occupies increasing fractions 
    of the available volume. In the second set of simulations, we varied the packing fraction 
    of the tracers while keeping the crowding polymer matrix at $\phi_P=0.1$. In both cases,
    we considered both kinds of tracer-tracer interactions, namely single-length scale (WCA) and 
    double-length scale (CS),  and progressively increased the strength of the tracer-polymer 
    interaction, $\epsilon_{AB}$, from zero.

%==================================================================================================    
    \subsection{Diffusion and structure of the tracer fluid at different polymer 
                packing fractions $\phi_P$}
    \label{sec:Diff1}
    
\begin{figure}[t!]
        %\centering
        \includegraphics[width=0.8\textwidth]{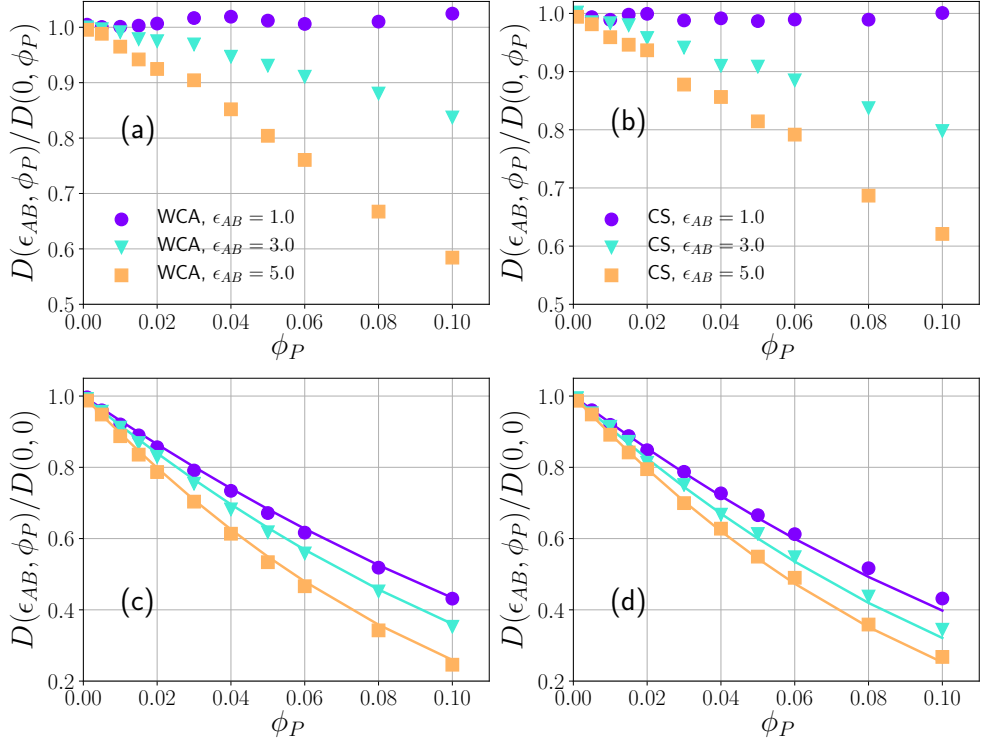}
        %\resizebox{0.5\linewidth}{!}{\input{DD0vsphip-tex.tex}}
        \caption{Diffusion coefficient of tracer molecules, $D(\epsilon_{AB}, \phi_P)$, as 
        a function of  the polymer 
        volume fraction $\phi_P$ for different values of the tracer-polymer affinity, $\epsilon_{AB}$.
        Left: WCA tracers. Right: CS tracers.
        Upper panels. The diffusion coefficients are normalized to the $\epsilon_{AB}=0$ value 
        at the same values of $\phi_P$ to highlight the effect of the tracer-polymer affinity.
        Lower panels. The diffusion coefficients are normalized to the value at 
        $\epsilon_{AB}=0, \phi_P=0$, to make the effect of varying both parameters explicit.
        Solid lines correspond to one-parameter fits to a 
        model of diffusion in porous media, where the polymer matrix 
        is modeled as an effective quenched suspension of hard spheres, see Eq.~\ref{e:Dpore}. 
        Points are the averages obtained from five independent runs. 
        Error bars are smaller than the size of symbols.
        }
        \label{fig3}
    \end{figure}

    The results of our simulations are illustrated in Fig.~\ref{fig3}, where we 
    chose two different ways to normalize the data.  The effect of the 
    tracer-polymer affinity $\epsilon_{AB}$ is best singled out by normalizing the 
    measured diffusion coefficient through $D(0, \phi_P)$. Each point in the plot 
    then quantifies the reduction in mobility caused by the tracer-crowder attractive interaction 
    irrespective of the additional reduction due to crowding, i.e. excluded-volume. 
    The first observation is that, surprisingly, a moderate attractive
    interaction seems to have no effect on the tracer mobility (circles in the upper panels).
    Increasing the strength of the tracer-polymer interaction beyond the value 
    $\epsilon_{AB}=\epsilon$, that is, the typical repulsive energy at the monomer-tracer
    contact distance, tracer mobility appears progressively more and more hindered.
    In other words, a crowded environment that is also somewhat {\em sticky} causes 
    more hindrance to tracer diffusion. This effect is larger the more crowded the 
    environment, with  $D(\epsilon,\phi_P)/D(\epsilon=0,\phi_P)$ appearing to 
    decrease linearly with the crowding packing fraction $\phi_P$. It is 
    interesting to note that, based on the findings reported in Ref.~\cite{Putzel2014},
    we might expect the slope of the straight lines 
    $D(\epsilon_{AB},\phi_P)/D(0,\phi_P)$ to depend non-monotonically on 
    the tracer-polymer interaction strength $\epsilon_{AB}$. However, to investigate 
    whether the interesting non-monotonicity reported for  
    the setting considered in Ref.~\cite{Putzel2014}
    would also show up in our system would require 
    exploring more finely spaced values of  $\epsilon_{AB}$, possibly 
    in the predicted interval $\epsilon_{AB} \in [1,3]$ $k_BT$.\\
    %
    %%%%%%%%%%%%%%%%%%%%%%%%%%%%%%%%%%%%%%%%%%%%%%%%%%%%%%%%%%%%%%%%%%%%%%%%%%%%%%%%%%%%%%%%%%%%%%%%%%%%%%%%%%%%        
    \setcounter{subfigure}{0}% Reset subfigure counter
%%%%%%%%%%%%%%%%%%%%%%%%%%%%%%%%%%%%%%%%%%%%%%%%%%%%%%%%%%%%%%%%%%%%%%%%%%%%%%%%%%%%%%%%%%%%%%%%%%%%%%%%%%%%        
    \indent The second observation is that the presence of multiple length scales in the tracer-tracer 
    interaction does not seem to induce noticeable differences in the way their mobility 
    is shaped by the interaction with the environment. 
    As a matter of fact, the reduction in diffusion for the WCA and CS fluid particles 
    are very similar. While this may come to little surprise, as the tracer volume 
    fraction is still on the low side ($\phi_f=0.1$), we shall see that this
    observation on mobility is not mirrored by the corresponding static structure of the
    tracer fluid, which is characterized by different spatial correlations induced 
    by the crowders that indeed appear to depend on the kind of tracer-tracer interaction.\\
    \indent The lower panels in Fig.~\ref{fig3} show the same data normalized in a different 
    fashion, intended to highlight the combined action of crowder-tracer affinity and 
    crowding volume fraction on diffusion. It can be clearly appreciated that an environment that 
    is both crowded and somewhat {\em sticky} induces a substantial slowing down of the 
    mobility. Tracers appear to loose between 60 and 70 \% of their mobility at a polymer 
    packing fraction as low as 10 \% when the tracer-polymer attractive energy 
    is between 3 to 5 times the typical energy $\epsilon$. Again, this normalization does 
    not reveal substantial differences in mobility ascribed to the kind of tracer-tracer
    interaction in the bulk.  \\
    %=========================================================================================
    % THEORETICAL MODELS FROM EXCESS ENTROPY ..... MORE COMPLICATED TO ADAPT 
    % TO OUR CASE THAN I THOUGHT ...
    %
    %\cite{Rosenfeld1999,Rosenfeld1977}
    %
    %\begin{equation}
    %D(\phi_P,\phi_f,\epsilon) = D(0,\phi_f,\epsilon) \,e^{a(\epsilon)S_{ex}(\phi_P,\phi_f)}
    %\end{equation}        
    %
    %\begin{equation}
    %\label{e:SexM}
    %-\frac{S_{ex}}{N k_B} = -\frac{3(1+y_3)}{2} + \frac{2y_3}{1-\phi} + 
    %                         \frac{3(1-y_3/3)}{2(1-\phi)^2}
    %                         +(y_3-1)\log(1-\phi)
    %\end{equation}    
    %
    %where $y_3 = \phi^{-1/3}$ and $\phi = \phi_P + \phi_f$ is the 
    %total packing fraction of the polymer-tracer mixture.
    %
    \indent It is instructive to compare our results with known  predictions 
    of diffusion in complex porous media. In such context,
    the ratio of the tracer diffusivity in the matrix to the 
    Stokes-Einstein diffusivity $D_0$
    in the pure suspending fluid is known as the so-called {\em tortuosity} $\tau$.
    For a static matrix of spherical particles of diameter $\sigma_1$ 
    with volume fraction $\phi$, one has~\cite{Peppin2019}
    \begin{equation}
    \label{e:Dpore}
    \frac{D(\phi)}{D_0} \equiv \tau(\phi) = (1-\phi)^\nu \left[ 
                                                       1-\lambda_p(\phi)
                                                    \right]^{a + b\lambda_p(\phi)}
    \end{equation}    
    where $\nu = 0.4$, $a = 4.2$, $b = 0.55$  and $\lambda_p(\phi)$
    is the ratio of tracer particle diameter $\sigma$ to the typical 
    pore diameter $\sigma_p(\phi)$,
    \begin{equation}
    \label{e:lambdap}
    \lambda_p(\phi) \equiv \frac{\sigma}{\sigma_p(\phi)} = 
    \left( \frac{\sigma}{\sigma_1} \right) \frac{3\phi}{1-\phi}
    \end{equation}    
    It is interesting to inquire whether our  polymer matrices behave effectively
    as simple quenched suspensions of hard spheres by treating the diameter 
    of such effective spheres $\sigma_1$ as an adjustable parameter. The fits shown in 
    the bottom panels in Fig.~\ref{fig3} reveal that our polymer matrices indeed 
    behave as quenched suspensions of hard spheres over the whole range of 
    parameters considered. More precisely, as the tracer-polymer interaction strength 
    increases, the typical effective pore size decreases, showing that {\em stickier}
    polymer matrices slow down tracers as quenched suspensions with smaller pores 
    would do. This analogy also allows us to compute the equivalent void percolation threshold 
    of the polymer matrix $\phi_P^c$, that is, the critical packing fraction
    where the pore size equals the tracer size. From the condition $\lambda_p=1$,
    Eq.~\eqref{e:lambdap} gives immediately
    \begin{equation}
    \label{e:phipc}
    \phi_P^c = \frac{\sigma_1}{3\sigma + \sigma_1}
    \end{equation}
    The best-fit values of the size of the effective crowding spheres 
    and the corresponding 
    percolation thresholds are reported in Table~\ref{tab: 1}. \\
%
%%%%%%%%%%%%%%%%%%%%%%%%%%%%%%%%%%%%%%%%%%%%%%%%%%%%%%%%%%%%%%%%%%%%%%%%%%%%%%%%%%%
    \begin{table}
        \centering
        \begin{tabular}{ r r r r r r r  } 
        \toprule
        \multicolumn{1}{c}{} & & \multicolumn{2}{c}{WCA} & & \multicolumn{2}{c}{CS} \\ 
        \cmidrule(lr){2-4} \cmidrule(lr){6-7}
        \multicolumn{1}{c}{$\epsilon_{AB}$}   & &
        \multicolumn{1}{c}{$\sigma_1/\sigma$} & 
        \multicolumn{1}{c}{$\phi_P^c$}        & &
        \multicolumn{1}{c}{$\sigma_1/\sigma$} & 
        \multicolumn{1}{c}{$\phi_P^c$}        \\ 
        \midrule
        0.0 & & $1.85  \pm  0.02$ & $0.381 \pm 0.004$ & & $1.89  \pm  0.02 $ & $0.386 \pm 0.002$\\         
        1.0 & & $1.91  \pm  0.02$ & $0.389 \pm 0.002$ & & $1.86  \pm  0.03 $ & $0.383 \pm 0.004$\\ 
        3.0 & & $1.57  \pm  0.01$ & $0.343 \pm 0.001$ & & $1.51  \pm  0.02 $ & $0.335 \pm 0.003$\\ 
        5.0 & & $1.21  \pm  0.01$ & $0.287 \pm 0.002$ & & $1.26  \pm  0.01 $ & $0.296 \pm 0.002$\\ 
         \hline
        \end{tabular}
    \caption{Best-fit values of the  diameter of the effective crowding hard spheres,
    $\sigma_1$, obtained by fitting  Eq.~\ref{e:Dpore} to our data over 
    the whole  range of polymer packing fraction $\phi_P$. The corresponding  
    void percolation tresholds $\phi_P^c$ computed from Eq.~\ref{e:phipc} are also reported.}
    \label{tab: 1}
    \end{table}
%%%%%%%%%%%%%%%%%%%%%%%%%%%%%%%%%%%%%%%%%%%%%%%%%%%%%%%%%%%%%%%%%%%%%%%%%%%%%%%%%%%
%
    \indent The similar trends observed in the mobility of WCA and CS particles
    might still conceal some more conspicuous difference in the structural 
    reorganization of the 
    tracer fluid induced by the sticky crowding matrices. 
    In order to investigate this aspect, 
    it is instructive to compute the radial distribution function (RDF), both for 
    tracer-tracer pairs (BB) as well as for tracer-monomer pairs (AB).    
%   
%%%%%%%%%%%%%%%%%%%%%%%%%%%%%%%%%%%%%%%%%%%%%%%%%%%%%%%%%%%%%%%%%%%%%%%%%%%%%%%%%%%%%%%%%%%%%%%%%%%%%%%%%%%%           
    \setcounter{subfigure}{0}% Reset subfigure counter
    \begin{figure}[h!]
        \includegraphics[width=16truecm]{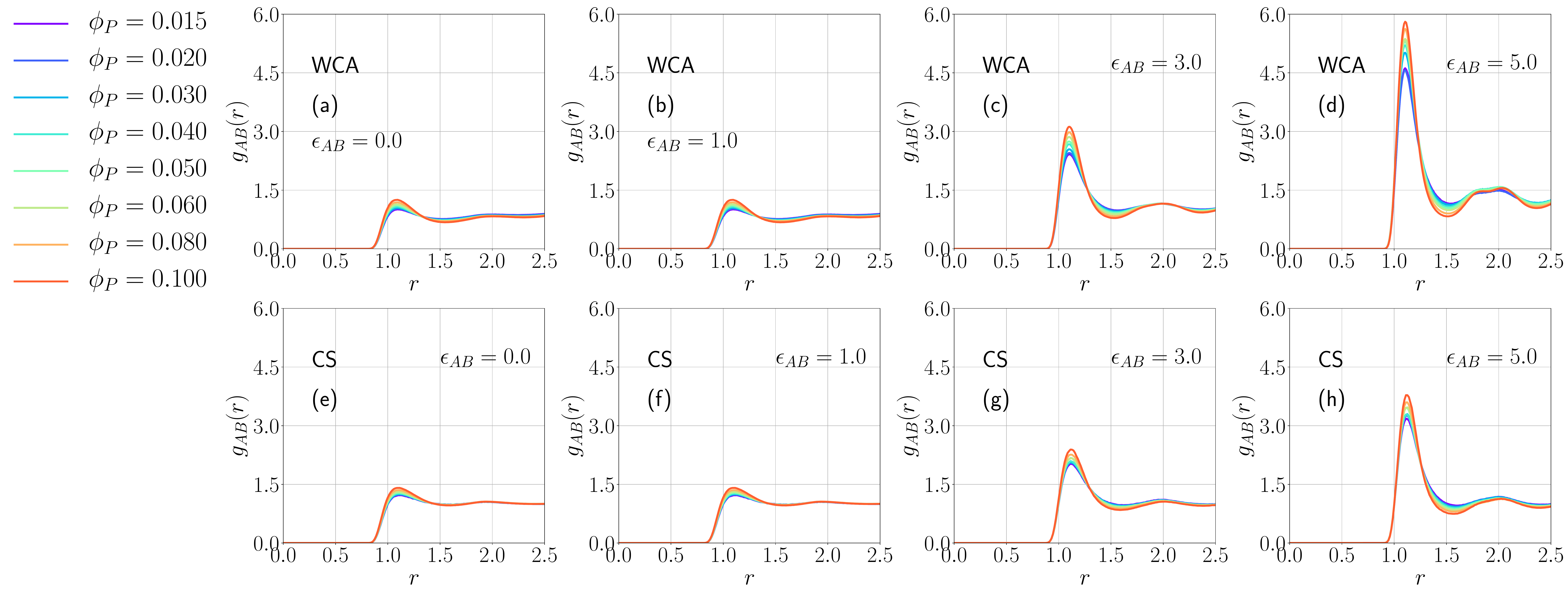} 
        %\subfigure[]{\includegraphics[width=0.30\textwidth]{rdfAB_WCA_eps1_fixed_fluid}}
        %\subfigure[]{\includegraphics[width=0.30\textwidth]{rdfAB_CS_eps1_fixed_fluid}}
        %\subfigure[]{\includegraphics[width=0.30\textwidth]{rdfAB_WCA_eps3_fixed_fluid}} 
        %\subfigure[]{\includegraphics[width=0.30\textwidth]{rdfAB_CS_eps3_fixed_fluid}}
        %\subfigure[]{\includegraphics[width=0.30\textwidth]{rdfAB_WCA_eps5_fixed_fluid}}
        %\subfigure[]{\includegraphics[width=0.30\textwidth]{rdfAB_CS_eps5_fixed_fluid}}
        \caption{The polymer-tracer RDF $g_{AB}(r)$ for WCA (top) and (CS) particles for 
        different values of the tracer-polymer affinity.}
        \label{fig4}
    \end{figure}
%%%%%%%%%%%%%%%%%%%%%%%%%%%%%%%%%%%%%%%%%%%%%%%%%%%%%%%%%%%%%%%%%%%%%%%%%%%%%%%%%%%%%%%%%%%%%%%%%%%%%%
%%%%%%%%%%%%%%%%%%%%%%%%%%%%%%%%%%%%%%%%%%%%%%%%%%%%%%%%%%%%%%%%%%%%%%%%%%%%%%%%%%%%%%%%%%%%%%%%%%%%%%    
    \setcounter{subfigure}{0}% Reset subfigure counter
    \begin{figure}[h!]
        \centering
        \includegraphics[width=16truecm]{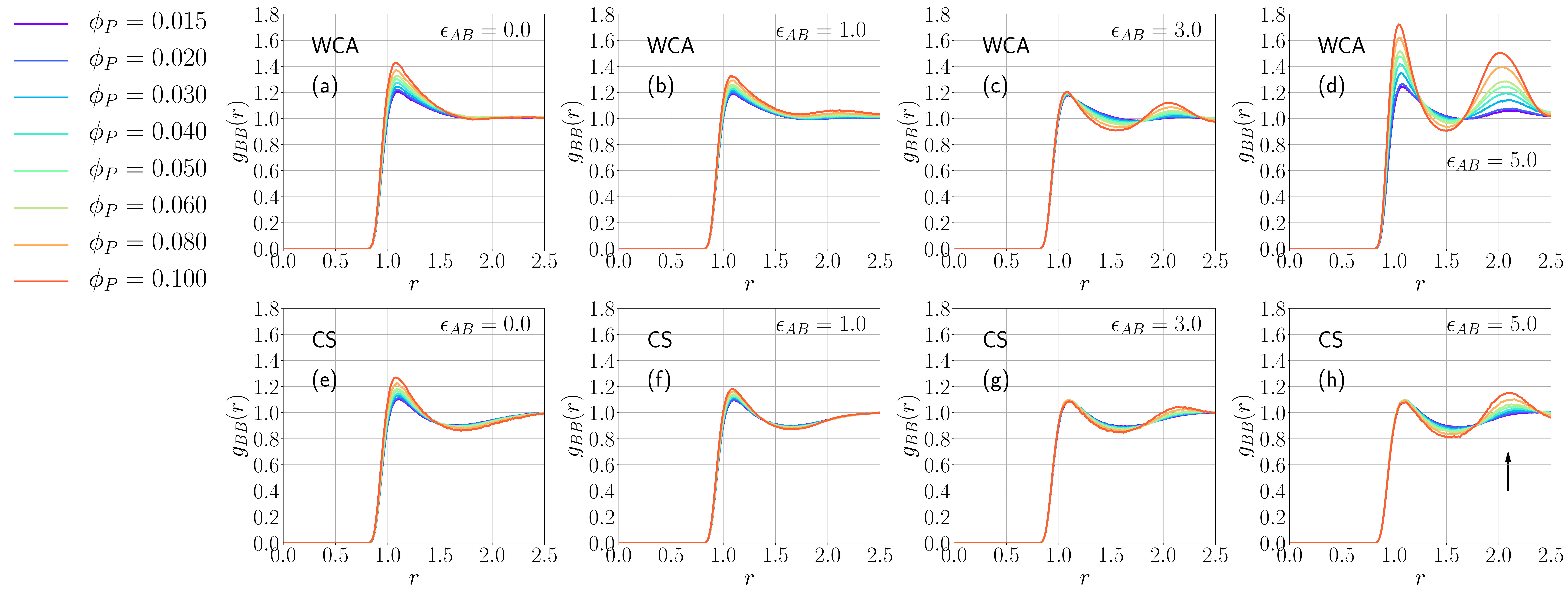}
        %\subfigure[]{\includegraphics[width=0.30\textwidth]{rdfBB_WCA_eps1_fixed_fluid}}
        %\subfigure[]{\includegraphics[width=0.30\textwidth]{rdfBB_CS_eps1_fixed_fluid}}
        %\subfigure[]{\includegraphics[width=0.30\textwidth]{rdfBB_WCA_eps3_fixed_fluid}}
        %\subfigure[]{\includegraphics[width=0.30\textwidth]{rdfBB_CS_eps3_fixed_fluid}}
        %\subfigure[]{\includegraphics[width=0.30\textwidth]{rdfBB_WCA_eps5_fixed_fluid}}
        %\subfigure[]{\includegraphics[width=0.30\textwidth]{rdfBB_CS_eps5_fixed_fluid}}
        \caption{The tracer-tracer RDF $g_{BB}(r)$ for WCA (top) and (CS) particles for 
        different values of the tracer-polymer affinity.}
        \label{fig5}
    \end{figure}
%%%%%%%%%%%%%%%%%%%%%%%%%%%%%%%%%%%%%%%%%%%%%%%%%%%%%%%%%%%%%%%%%%%%%%%%%%%%%%%%%%%%%%%%%%%%%%%%%%%%%%%%%%%%            
%              
    Figure~\ref{fig4} illustrates the behavior of the polymer-fluid RDF,   $g_{AB}(r)$.
    Both kind of fluids appear to develop an increasing degree of structural organization 
    in the vicinity of the polymer matrix
    as the polymer-tracer affinity increases beyond $\epsilon_{AB} =1$. However, this 
    analysis reveals that CS particles are much less prone to affinity-induced structural 
    organization around the crowders. This observation can be interpreted as a direct result 
    of the competition between different length scales in the repulsion between CS particles,
    which appears to induce some frustration in the spatial ordering of such tracers
    in the presence of crowding.\\
    \indent A direct inspection of the tracer-tracer RDFs $g_{BB}(r)$ (Fig.~\ref{fig5}) 
    confirms that structural ordering is globally more hindered in the CS fluid. 
    Interestingly,
    WCA particles get more and more structured around each other at large values of 
    $\epsilon_{AB}$ and 
    the first and second coordination shells appear to become populated at 
    essentially the same rate (see panel (d) in Fig.~\ref{fig5}).
    This is most likely a {\em templating} effect, whereby the chain connectivity 
    of the sticky polymer matrix causes ordered association of the tracers 
    along the polymer chains, thus promoting the development of 
    structural correlations in the vicinity of the crowders. This 
    effect would most likely disappear in the case of crowding agents 
    whose micro-structure would be incommensurate to that of the tracers.  
    We will come back to this effect in the next section. 
    By contrast, the dual length-scale repulsion of CS particles appears to 
    essentially suppress any appreciable structuring in the first coordination 
    shell, while the second shell becomes more and more populated, even if to 
    a much lesser extent when compared to WCA particles  (see arrow in 
    panel (h) in Fig.~\ref{fig5}).\\
    %
%================================================================================================
    \subsection{The effect of self-crowding on diffusion}
    \label{sec:Diff2}

    It is interesting to reverse the diffusion analysis illustrated above, where the
    fluid packing fraction was held fixed at $\phi_f=0.1$, while we investigated  
    the combined effect of $\epsilon_{AB}$ and the crowding volume occupancy, $\phi_P$.
    In this section we analyze the results of simulations where the crowding density was 
    fixed at the same moderate value, i.e. $\phi_P=0.1$, 
    while we increased progressively self-crowding by  
    letting $\phi_f$ and $\epsilon_{AB}$ vary.
%    
%%%%%%%%%%%%%%%%%%%%%%%%%%%%%%%%%%%%%%%%%%%%%%%%%%%%%%%%%%%%%%%%%%%%%%%%%%%%%%%%%%%%%%%%%%%%%%%%%%
    \begin{figure}[t!]
        \centering 
        \includegraphics[width=0.5\textwidth]{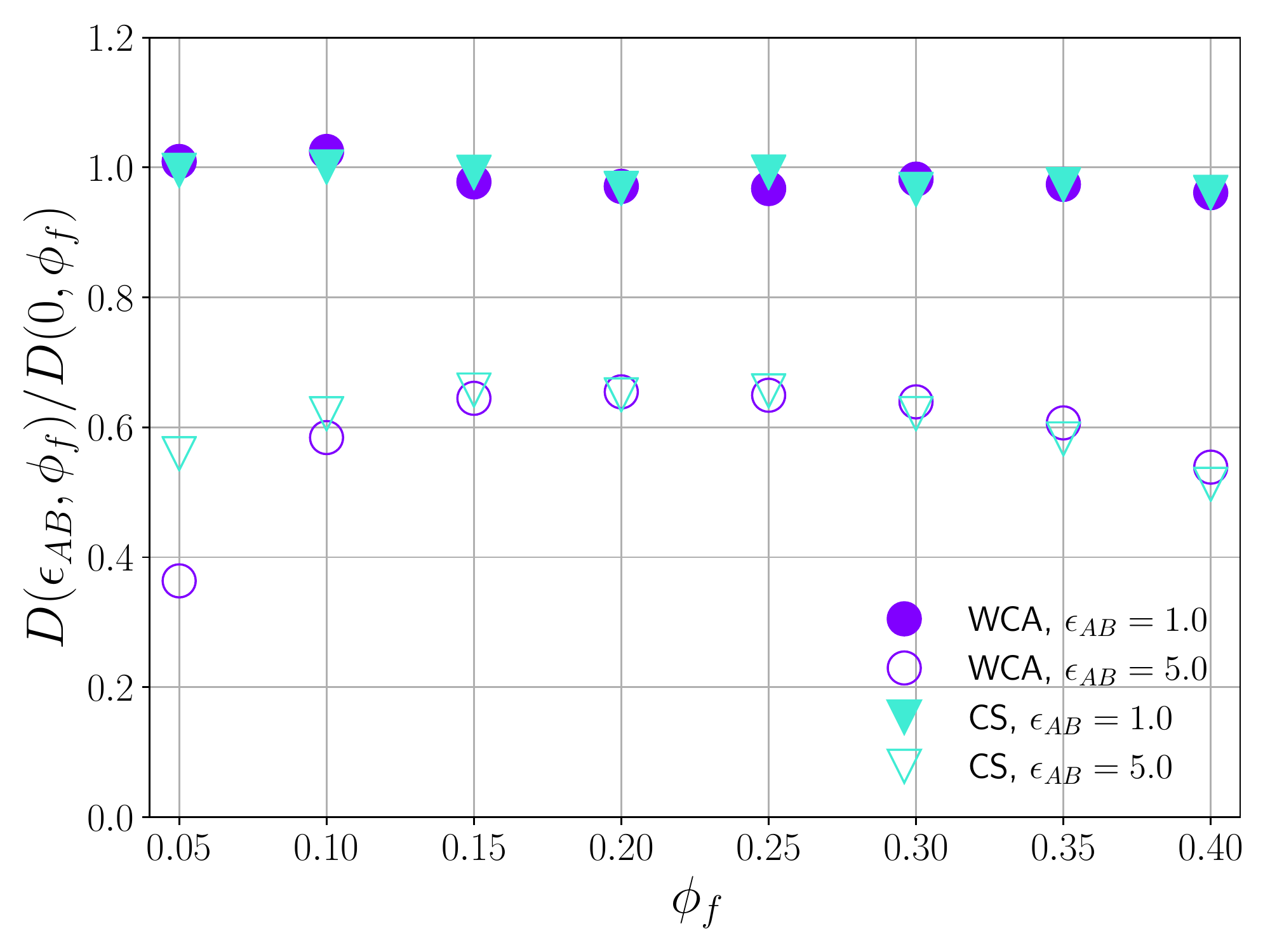}
        \caption{Diffusion coefficient of tracers normalized to the purely repulsive 
        values $D(0, \phi_f)$ as a function of the fluid volume 
        fraction $\phi_f$ for two values of $\epsilon_{AB}$.}
        \label{fig6}
    \end{figure}
    \setcounter{subfigure}{0}% Reset subfigure counter
%%%%%%%%%%%%%%%%%%%%%%%%%%%%%%%%%%%%%%%%%%%%%%%%%%%%%%%%%%%%%%%%%%%%%%%%%%%%%%%%%%%%%%%%%%%%%%%%%%%%%%%%%     
%
        In order to isolate the effect of the polymer-tracer affinity, we plot in figure~\ref{fig6}
        the diffusion coefficient of tracers $D(\epsilon_{AB}, \phi_f)$ 
        normalized to the purely repulsive value at the same density,  $D(0, \phi_f)$.
        As a first observation, it can be clearly appreciated  that varying the 
        fluid packing fraction (self-crowding)  seems to have a less dramatic effect on diffusion than 
        varying the crowding density.
        For low affinity, $\epsilon_{AB} = 1.0$,  WCA and the CS tracers
        display an approximately constant diffusion coefficient, mirroring the corresponding 
        trend observed in the reversed situation (see Fig.~\ref{fig3}, top panels).\\
        \indent An interesting phenomenon is observed in the high-affinity case.
        When $\epsilon_{AB} = 5.0$, both tracer species display a 
        water-like diffusion anomaly, more pronounced in the case of WCA tracers.
        It is clear from figure~\ref{fig6}
        that the diffusion constant increases as $\phi_f$ increase, reaches a 
        maximum at $\phi_f\cong 0.2 $ and then decreases again as
        the fluid density increases further. While for CS molecules
        this anomalous behavior is  well 
        known and directly ascribed to the dual-length 
        repulsion~\cite{barros2006thermodynamic,bordin2018},  it appears rather 
        unexpected in the simple repulsive WCA fluid.\\
%         
%%%%%%%%%%%%%%%%%%%%%%%%%%%%%%%%%%%%%%%%%%%%%%%%%%%%%%%%%%%%%%%%%%%%%%%%%%%%%%%%%%%%%%%%%%%%%%%%%%%%%%%%
    \setcounter{subfigure}{0}% Reset subfigure counter  
    \begin{figure}[t!]
        \centering
        \subfigure[]{\includegraphics[width=0.4\textwidth]{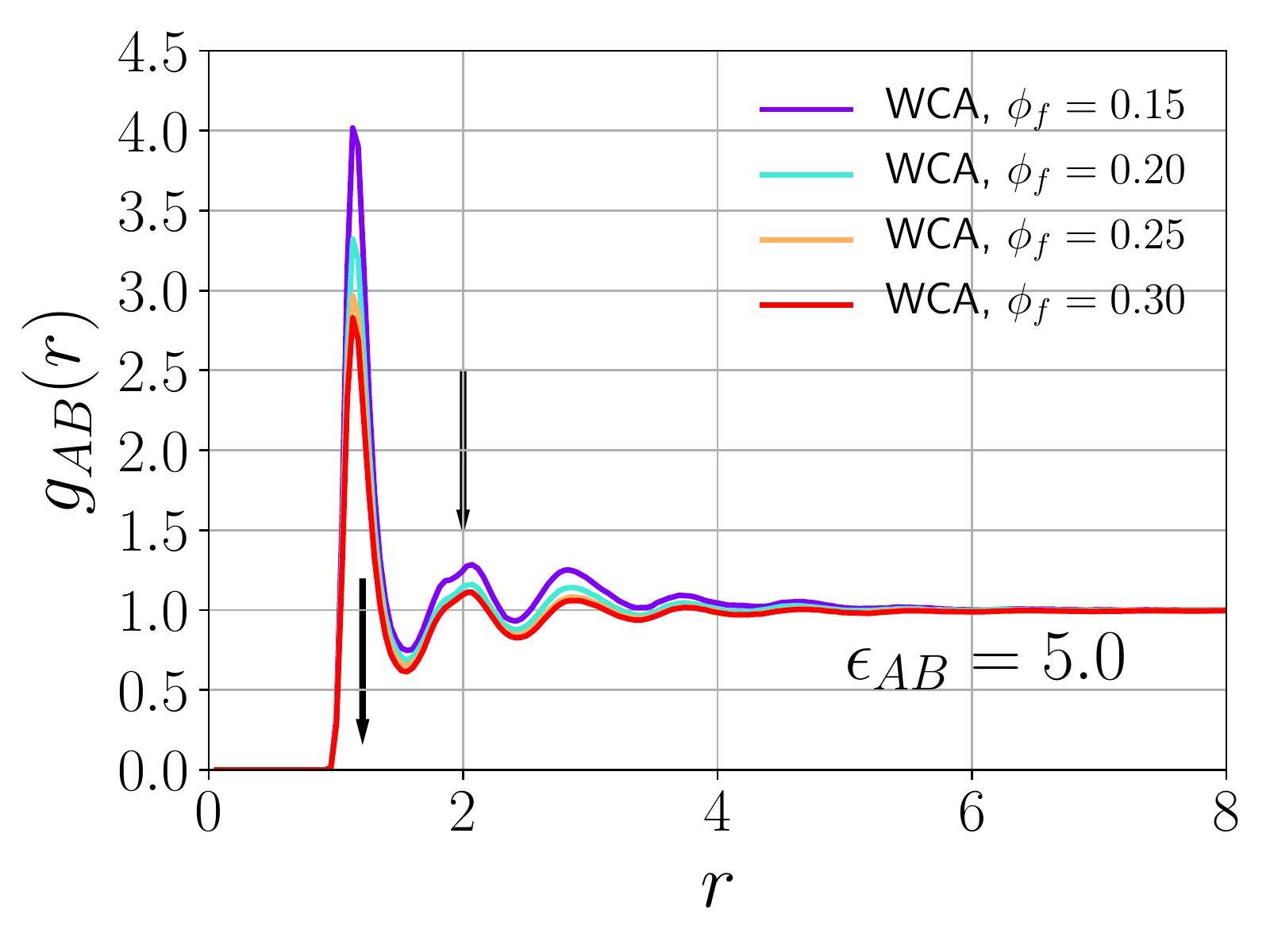}}
        \subfigure[]{\includegraphics[width=0.4\textwidth]{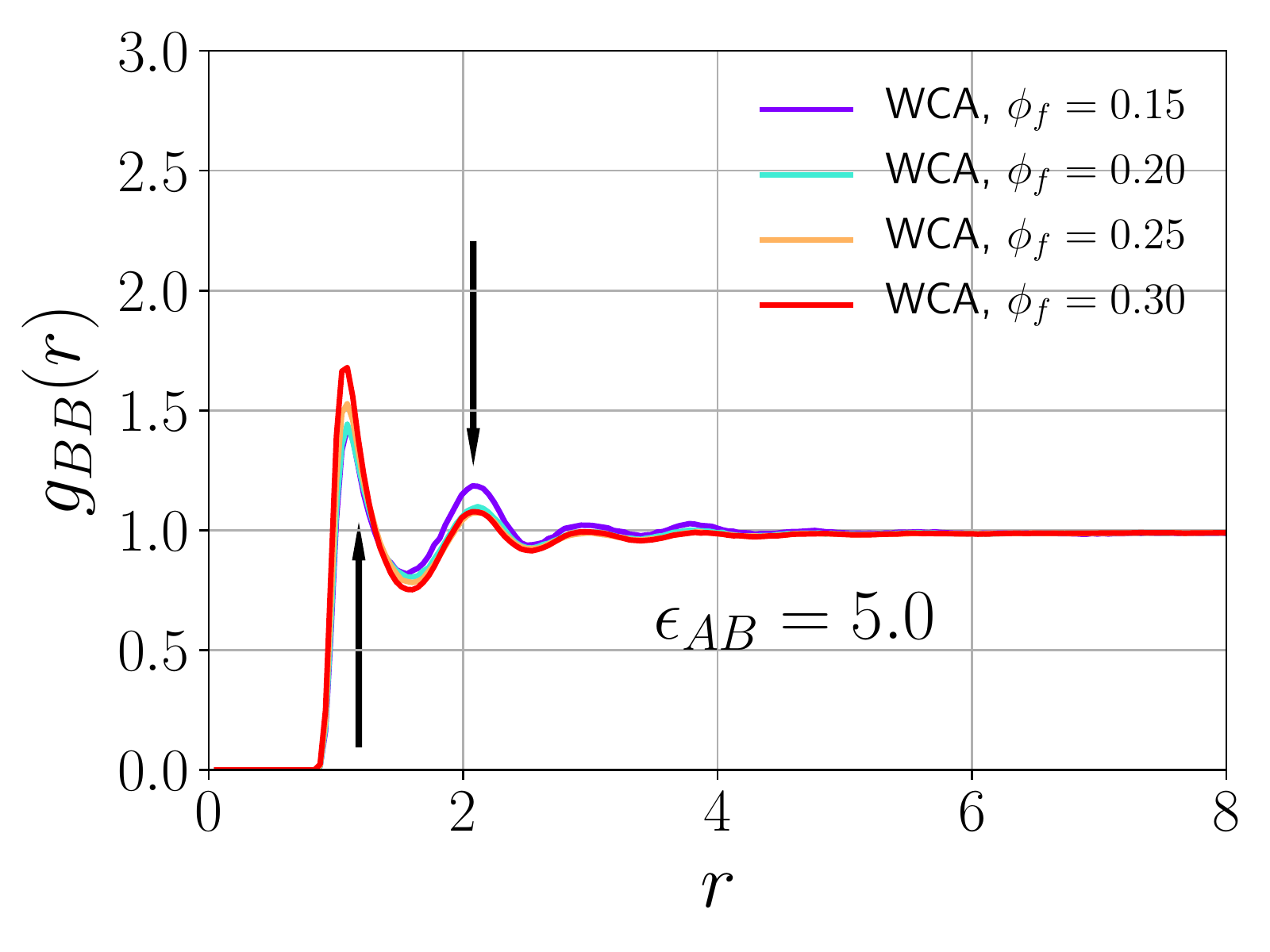}}
        \subfigure[]{\includegraphics[width=0.4\textwidth]{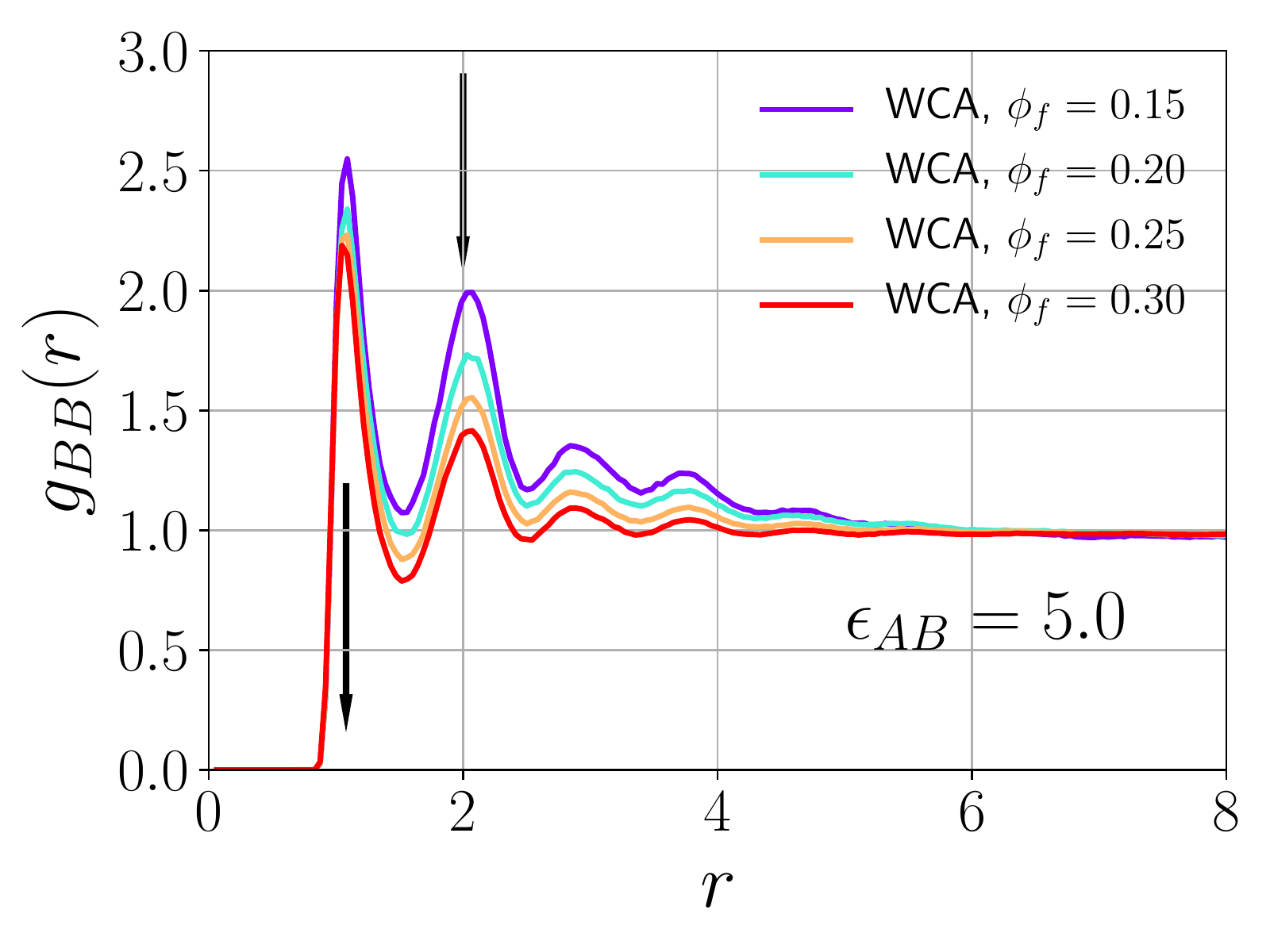}}
        \subfigure[]{\includegraphics[width=0.4\textwidth]{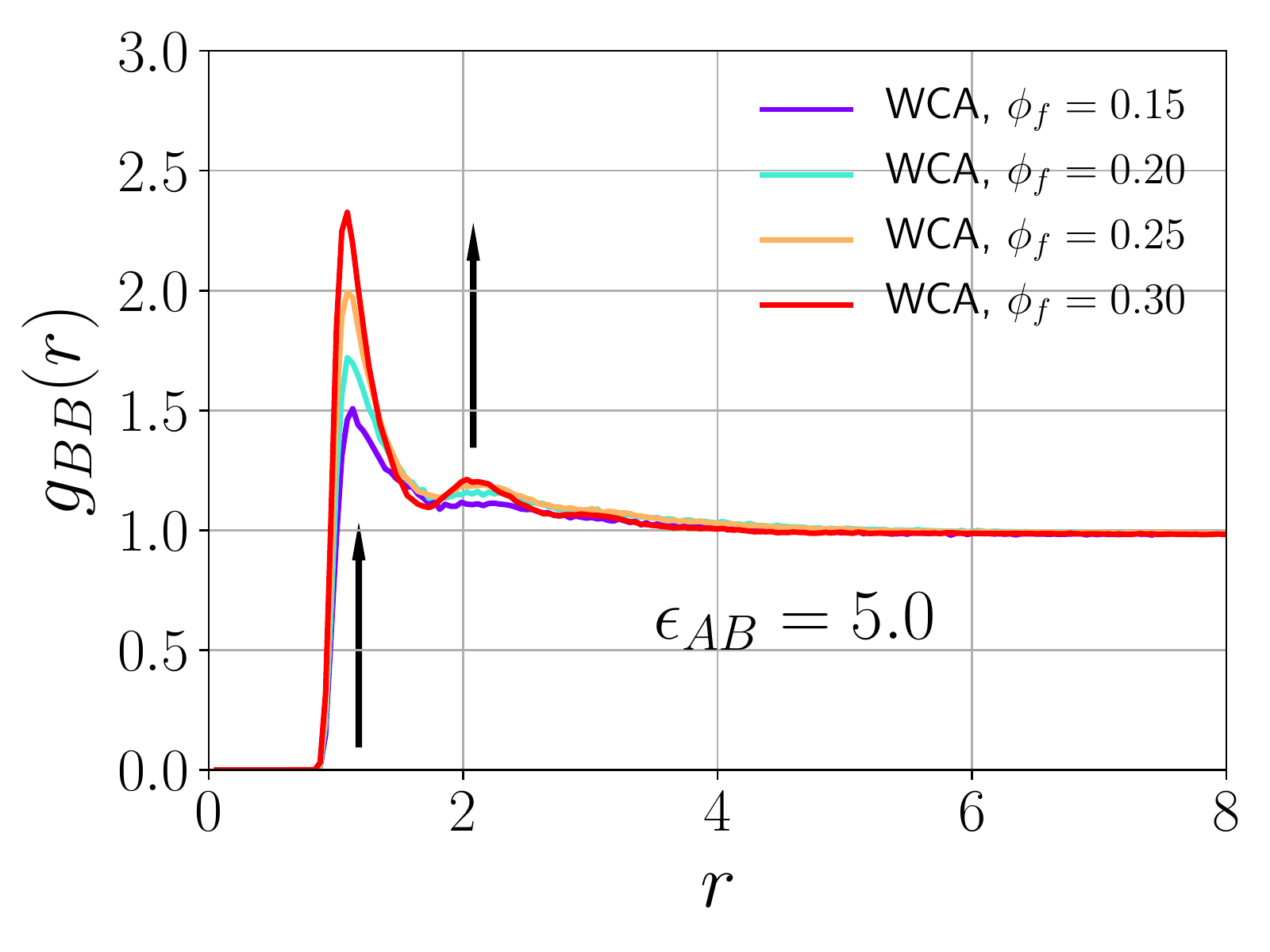}}
         \caption{Total (adsorbed + bulk) polymer-tracer (a) and tracer-tracer (b) 
         radial distribution functions at increasing values of the tracer packing fraction.
         Bottom panels: the tracer-tracer RDF for adsorbed tracers (c) and bulk tracers (d).
         The polymer matrix packing fraction is $\phi_P=0.1$.}
         \label{fig7}
     \end{figure}
%%%%%%%%%%%%%%%%%%%%%%%%%%%%%%%%%%%%%%%%%%%%%%%%%%%%%%%%%%%%%%%%%%%%%%%%%%%%%%%%%%%%%%%%%%%%%%%%%%%%%%%%
%
        \indent Some insight into this unexpected result can again be gathered 
        by looking at the radial distribution functions at increasing values 
        of $\phi_f$ in the high-affinity case for WCA tracers (fig.~\ref{fig7}). 
        More precisely, one notices that the effect of increasing the
        fluid density is to progressively lower the fraction of tracers 
        adsorbed on the polymer matrix. 
        This behavior is clearly illustrated by the decrease of both the 
        first and second peaks in the polymer-tracer RDF, $g_{AB}(r)$ (panel (a)). 
        At the same 
        time, while the polymer-tracer interface becomes less organized, the tracers 
        start developing more short-range order. This is reflected by the increase 
        of the first peak of the 
        tracer-tracer RDF, $g_{BB}(r)$ (panel (b)).
        To confirm this, we divided the tracers in two classes: adsorbed and bulk tracers. 
        To this end, 1000 system snapshots were analyzed based on the inter tracer-polymer bonding, 
        in a similar manner as done in previous works to characterize colloids 
        aggregates~\cite{Toledano09, bordin2018b, Bordin19}. 
        Tracers were considered adsorbed when they lie at distances smaller than 1.25$\sigma$  
        from one polymer monomer. If the distance was larger than this threshold, 
        the tracer was considered in the bulk. 
        The adsorbed and bulk tracer-tracer RDFs are shown in Fig.~\ref{fig7}
        (c) and (d), respectively. It can be appreciated that the increase in the first 
        peak of the total tracer-tracer RDF, Fig.~\ref{fig7}(b), comes from the contribution of 
        bulk tracers, Fig.~\ref{fig7}(d) - both grow as the density increases. At the same time, this analysis 
        provides a visual and quantitative rationalization of the aforementioned
        {\em templating} effect, which manifests itself in increased 
        tracer-tracer organization at the interface at large values of 
        $\epsilon_{AB}$. 
        It can be appreciated that such effective attractive tracer-tracer 
        interaction appears to be rather long-range, extending over about 4 tracer diameters.
        This effect can be observed in Fig.~\ref{fig7}(c): 
        the peaks decay as $\phi_f$ increases. Furthermore, we see here clearly 
        that self-crowding has a disruptive 
        effect on the templating action exerted by the polymer matrix on tracers, 
        increasingly promoting structural order of the tracer fluid in the bulk (see again panel (d)).\\
        \indent To further clarify how the polymer {\em stickiness}  
        modulates the effect of self-crowding, we 
        evaluated the MSD of adsorbed and bulk tracers (Fig.~\ref{fig8}).
        In the case of low affinity, there are appears to be no apparent difference 
        between adsorbed and bulk tracers -- both species 
        diffuse similarly regardless of the fluid density (Fig.~\ref{fig8}(a), (b) and (c)). 
        However, at higher affinity, it is clear that at low tracer densities the 
        adsorbed tracers display a reduced mobility with respect to 
        bulk tracers (Fig.~\ref{fig8} (d)). Increasing self-crowding,
        the two species recover the same mobility, starting at a value around 
        $\phi_f = 0.20$ -- 
        the point where the tracer diffusion coefficient has a maximum (see
        again Fig.~\ref{fig6}). 
        The different evolution of the weight associated with bulk and adsorbed tracers 
        with the overall tracer density  is presumably the reason behind
        the non-monotonic trend observed in Fig.~\ref{fig6}.
%         
%%%%%%%%%%%%%%%%%%%%%%%%%%%%%%%%%%%%%%%%%%%%%%%%%%%%%%%%%%%%%%%%%%%%%%%%%%%%%%%%%%%%%%%%%%%%%%%%%%%%%%%%
        \setcounter{subfigure}{0}% Reset subfigure counter  
        \begin{figure}[h!]
        \centering
        \subfigure[]{\includegraphics[width=0.3\textwidth]{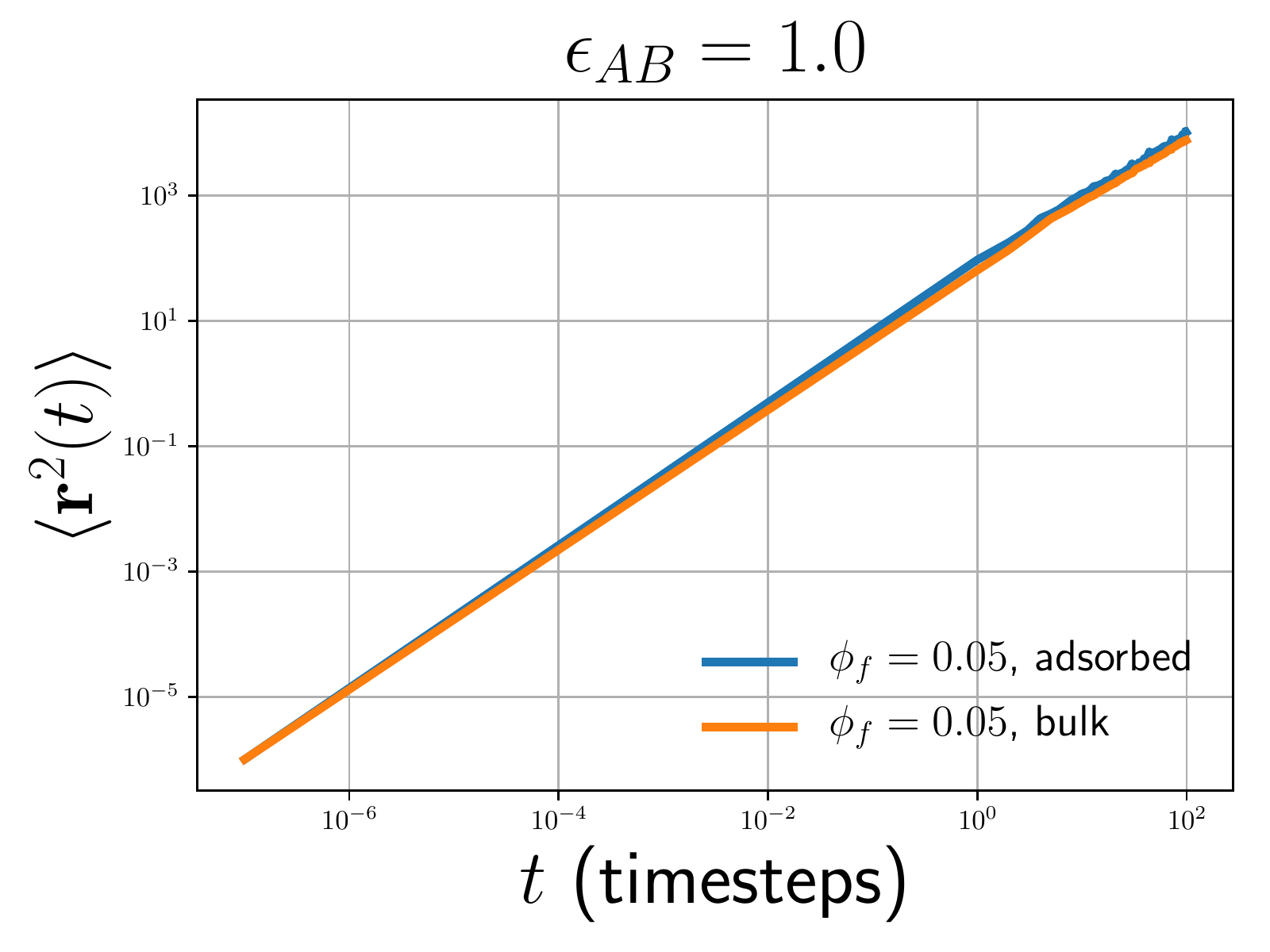}}
        \subfigure[]{\includegraphics[width=0.3\textwidth]{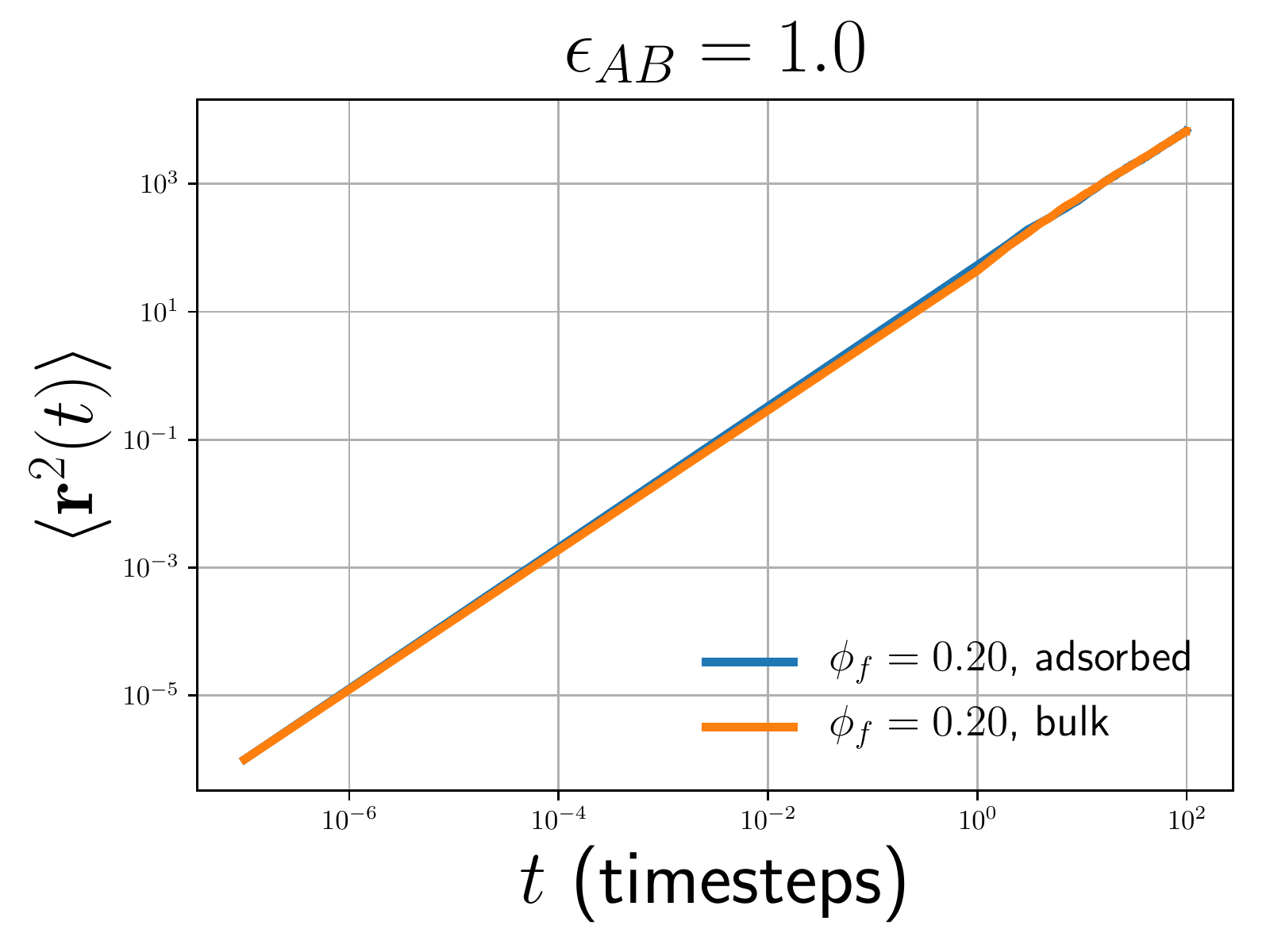}}
        \subfigure[]{\includegraphics[width=0.3\textwidth]{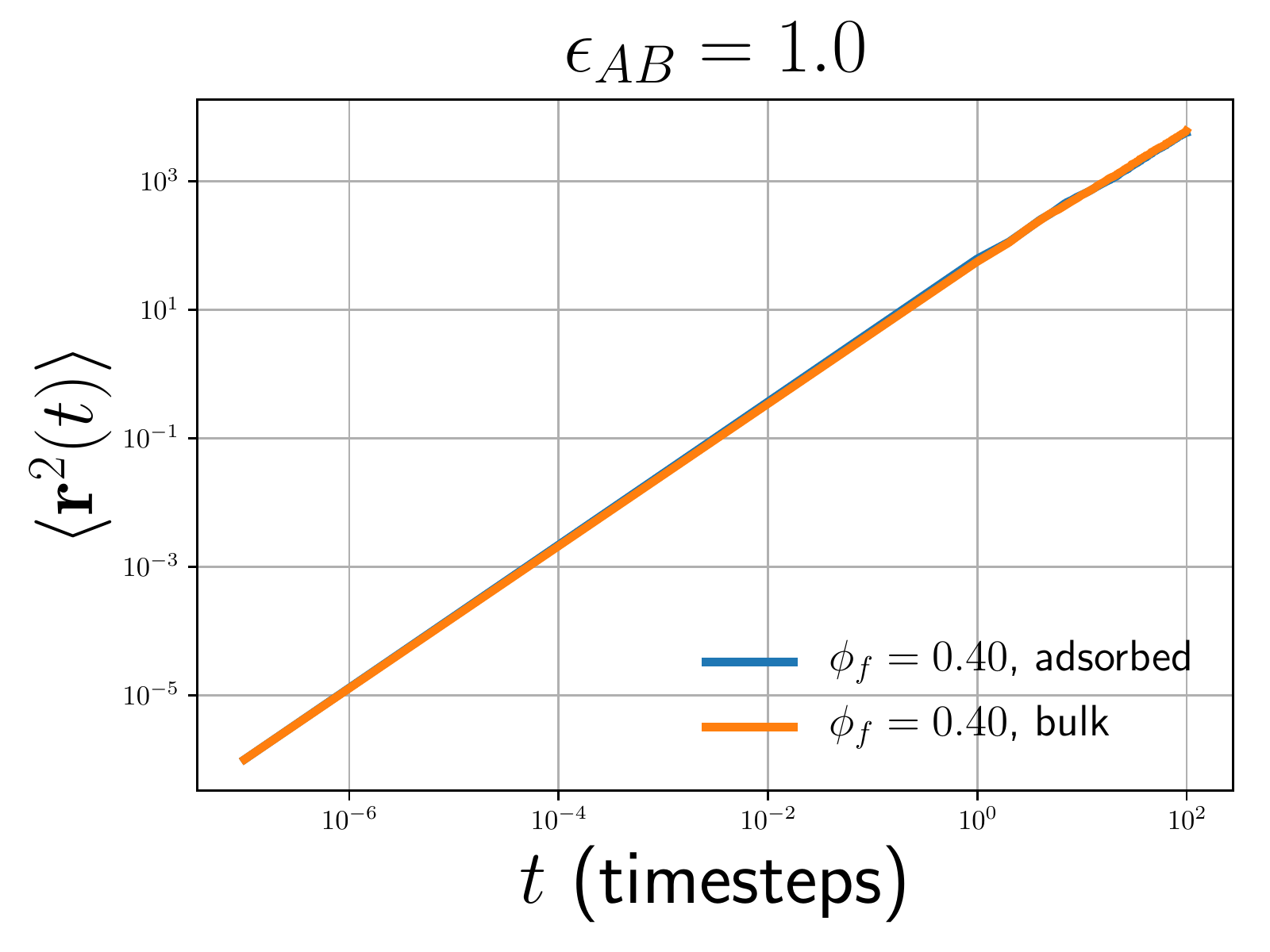}}
        \subfigure[]{\includegraphics[width=0.3\textwidth]{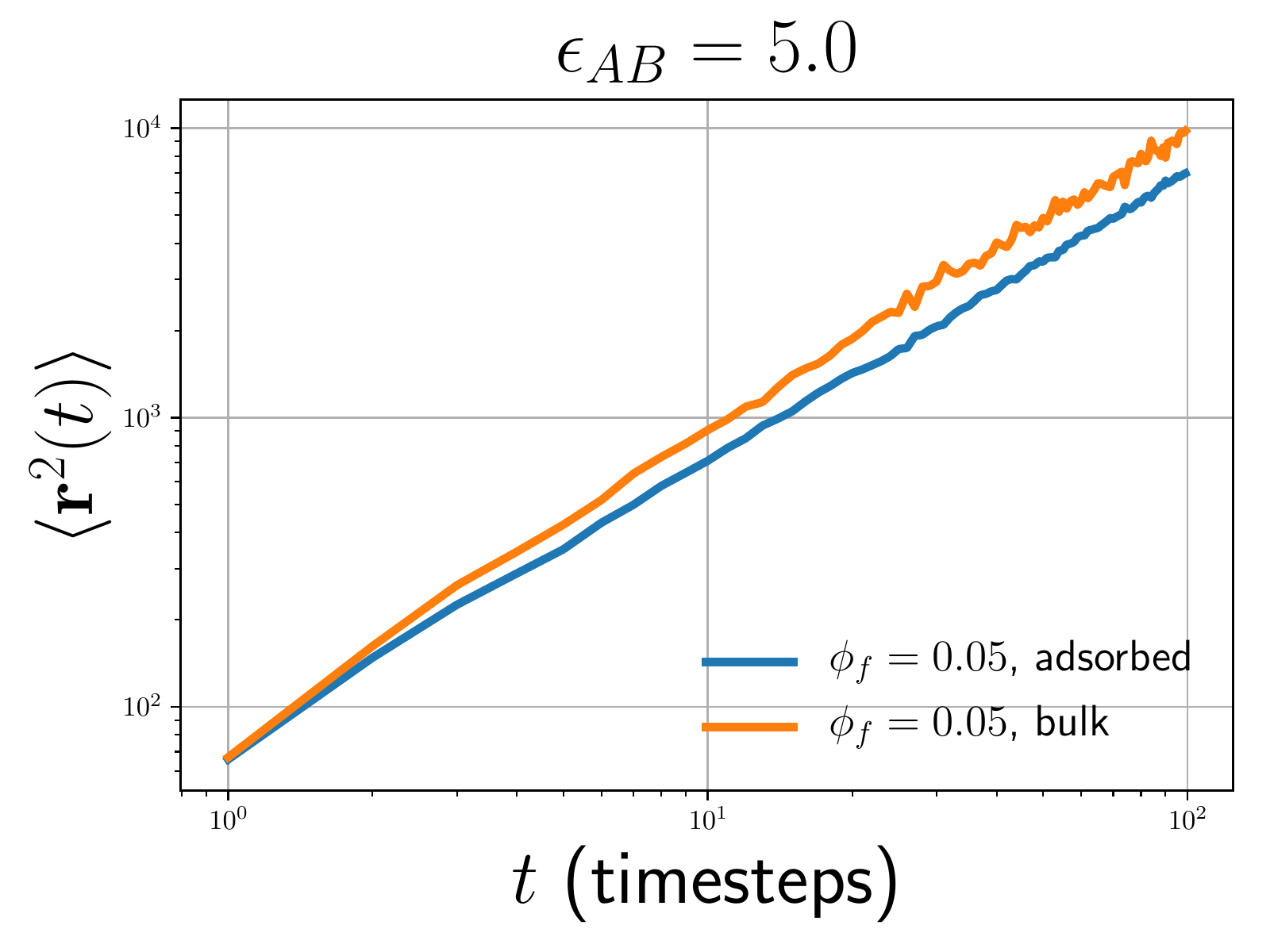}}
        \subfigure[]{\includegraphics[width=0.3\textwidth]{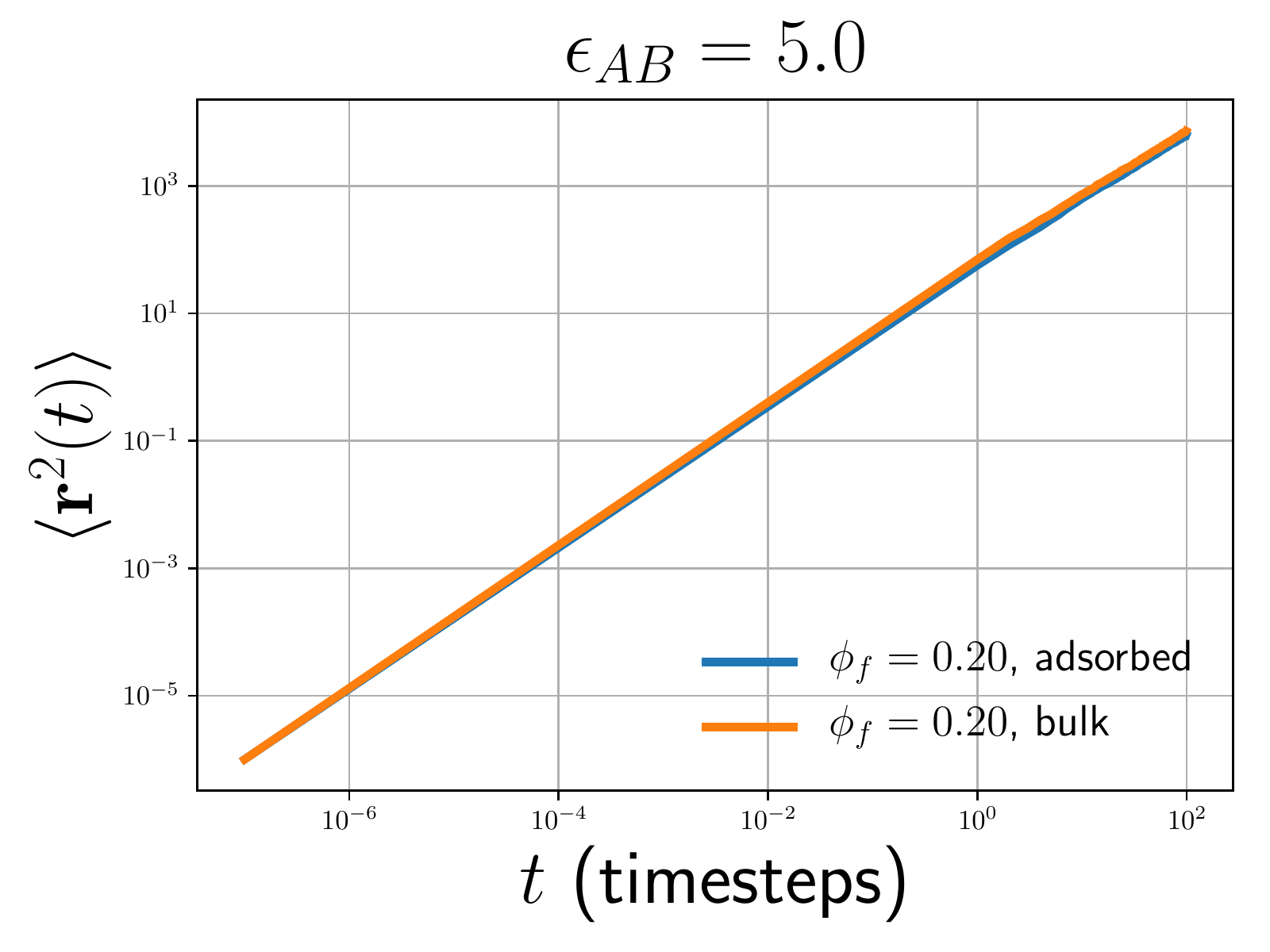}}
        \subfigure[]{\includegraphics[width=0.3\textwidth]{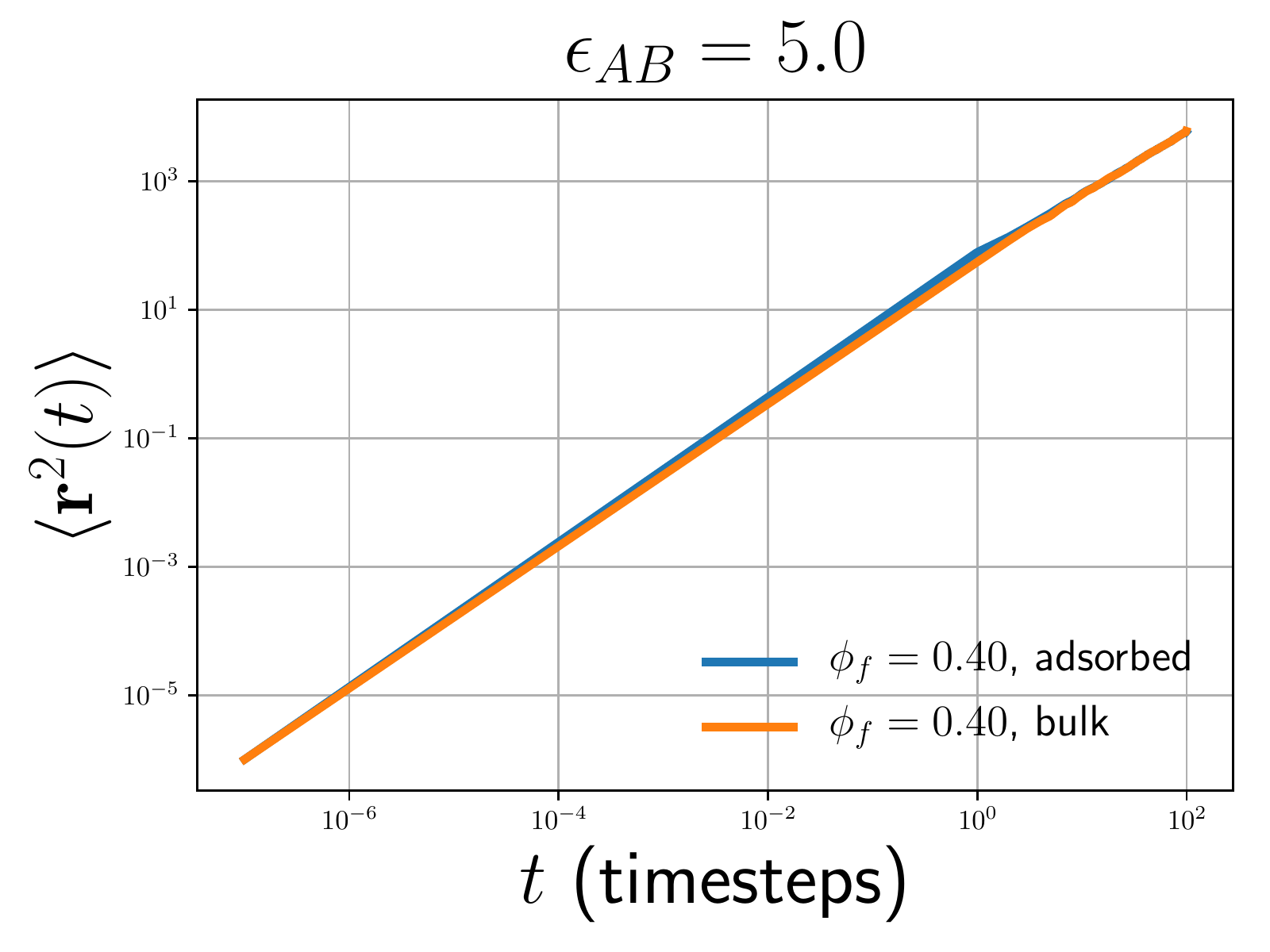}}
         \caption{MSD for adsorbed and bulk tracers. The top panels illustrate 
         the low-affinity case, while the bottom ones depict the  high-affinity case. 
         The fluid densities are: 0.05 (a) and (d); 0.20 (b) and (e); 0.40 (c) and (f). 
         The polymer matrix packing fraction is $\phi_P=0.1$.}
         \label{fig8}
     \end{figure}
%%%%%%%%%%%%%%%%%%%%%%%%%%%%%%%%%%%%%%%%%%%%%%%%%%%%%%%%%%%%%%%%%%%%%%%%%%%%%%%%%%%%%%%%%%%%%%%%%%%%%%%%
%
        In other words, as we increase
        the fluid density,  less and less tracers are adsorbed 
        by the polymer matrix, thereby raising the number of unconstrained, 
        fully mobile  molecules in the bulk. However, increasing the fluid density further, 
        the self-crowding effects become more prominent and the mobility of 
        bulk tracers starts decreasing
        as the fluid becomes more and more structured (see also the black arrows in Fig.~\ref{fig7}). Therefore, the combination of highly sticky obstacles and high 
        tracers density can lead to unexpected dynamical behaviors.
        
%==========================================================================================================     
    \section{Conclusions}
    
    In this paper, we employed large-scale Langevin dynamics simulations 
    to investigate the mobility  of tracer particles diffusing in a static matrix 
    consisting of quenched polymer chains of equal-sized monomers. 
    The two main parameters varied were the volume fraction occupied by 
    the polymer matrix and the strength of a short-range, non-specific 
    attractive interaction causing tracer particles to spend longer time 
    in the vicinity of the crowders.
    Furthermore, we focused on two types of tracer particles, characterized
    by different tracer-tracer repulsive interactions. In particular, we 
    considered core-softened tracers (CS), characterized by a dual-length repulsive 
    potential as compared to purely repulsive, shifted Lennard-Jones particles (WCA).\\
    \indent We found that excluded-volume interactions, i.e. crowding, reduce
    the tracer diffusion coefficient, all the more so the larger the polymer-tracer 
    affinity. At the moderate tracer volume fraction considered in this paper ($\phi_f=0.1$), 
    we found no appreciable signature of the kind of tracer-tracer interactions in 
    the measured diffusivity. The mobility of CS and WCA particles decreased with 
    increasing crowding and increasing affinity following practically indistinguishable 
    trends. In particular, we found that, for all the values of the attractive energy strength 
    considered, the {\em sticky} crowding matrices behaved as porous media consisting of 
    effective quenched suspensions of 
    purely repulsive hard spheres, as gauged by fitting a tortuosity model to our data.
    \indent If the measured diffusion coefficients bore no blueprint of the underlying 
    tracer-tracer interactions, the same was not true for static spatial correlations. 
    In particular, for large crowding tracer affinity, the core-softened tracers  showed 
    a considerably lower propensity to structure around the polymers, whereas WCA 
    particles showed substantial short- and intermediate-range order, increasing
    with the volume fraction of crowders. \\
    \indent  In order to explore further the distinctive signature of the tracer-tracer repulsion in the high affinity case,
    we ran a series of simulations at increasing density of tracers
    and intermediate crowding ($\phi_P=0.1$) for $\epsilon_{AB} = 5$,
    comparing the measured diffusion coefficient to the zero affinity case.
    While we recovered the known 
    water-like anomaly for CS particles, i.e. a non-monotonic trend of the diffusion 
    coefficient as a function of the tracer density for high affinity, we found 
    an even more pronounced anomaly of the same kind for the WCA particles.
    This kind of anomaly, which does not seem to be directly ascribed to  
    dual-length repulsion, is more deeply, and likely more generally rooted 
    in the competition between
    the confinement and the attraction exerted on tracers by the polymeric network.
    For dilute tracer fluids, an  increase in self-crowding is seen to induce 
    a progressive desorption of tracers from the polymer matrix, while the fluid
    gets more and more structured in the bulk. The apparent result of this is 
    that a little increase in $\phi_f$ from the very dilute case reduces 
    $D(\epsilon_{AB}=5,\phi_f)$ {\em less} than what observed in the 
    zero-affinity case, i.e. less than $D(\epsilon_{AB}=0,\phi_f)$.
    It is worth recalling that a relatively ample body of 
    literature exists reporting a non-monotonic 
    trend of the diffusion coefficient in the presence of crowding 
    and tracer-crowder attractive interactions as the strength of the latter 
    is increased. This scenario, also rooted in the competition between confinement and attraction, 
    appears rather general, from quenched-annealed mixtures 
    of hard spheres~\cite{Putzel2014}, to ions in a charged polymer gel~\cite{holmes1990},
    nanoparticles in polymer melts\cite{Yamamoto2011}
    ans simple hard-sphere like colloids~\cite{Pham104}. However, despite the large body of 
    work in this area, further work seems to be needed
    to gather a more comprehensive picture of tracer diffusion in the presence of 
    crowded and attractive media in the regime
    where self-crowding effects become important.\\
    \indent As a final remark, we point out that no evidence of anomalous diffusion was ever
    found in the results of our simulations. 
    In this kind of systems,
one could expect to observe (at least) two kinds of anomalous diffusion in some regimes, (i)
transient anomalous behaviour as a cross-over between short-time diffusion 
and long-time diffusion and (ii) (possibly asymptotic) anomalous diffusion of the single-file kind
near the void percolation threshold. The former case would flag the presence of some caging effect,
whereby a quick diffusive exploration of the local cage is followed by a much slower 
diffusion over the large-scale ensemble of cages. We did not find any evidence of a transition 
between short-time  and long-time diffusion for any of the parameter choices considered 
in this work.  The reason is likely to be related to the specific way the space where 
tracers diffuse is crowded with obstacles. Since these are self-avoiding polymer {\em chains}, 
the presence of cages at moderate packing fraction (we considered up to $\phi_P=0.1$)
should be negligible. It would be interesting to investigate how tracers diffuse within 
much denser crowding matrices 
in order to explore whether caging effects emerge at all in the presence of connected 
crowders. Regime (ii) is a subtler matter altogether. 
While some evidence of anomalous diffusion has been reported in the case of quenched-annealed 
mixtures of hard spheres~\cite{Kurzidim2011,Kurzidim2010}, it is uncertain whether this would be 
a general feature of obstructed diffusion.  
In the present work, we did not explore the
regime close to the void percolation threshold, neither we performed quantitative characterizations
of the void size distributions. It would be extremely  interesting to couple these kinds of  studies
to measurements of tracer mobility, to investigate 
whether anomalous diffusion related to a phase transition of the void percolation type 
emerges in the presence of connected obstacles such as immobile polymer chains.

    \section{Acknowledgment}

  This study was financed in part by the Coordena\c{c}\~ao de Aperfei\c{c}oamento de Pessoal de N\'ivel Superior (CAPES), Finance Code 001.
  JRB acknowledge the Brazilian agencies CNPq and FAPERGS for financial support. 
  FP is greatly indebted to Giuseppe Foffi for illuminating discussions.

\bibliography{ref}

\end{document}